\documentclass{emulateapj}
\usepackage{graphics}
\usepackage{lscape}

\slugcomment{Accepted to \textit{The Astrophysical Journal}}

\newcommand{\chandra}{\textit{Chandra }}
\newcommand{\spitzer}{\textit{Spitzer }}

\begin{document}


\shortauthors{Temim et al.}

\shorttitle{``G54.1+0.3 with Chandra, Spitzer''}

\title{Deep Chandra Observations of the Crab-like Pulsar Wind Nebula G54.1+0.3 and Spitzer Spectroscopy of the Associated Infrared Shell}

\author{TEA TEMIM\altaffilmark{1,2}, PATRICK SLANE\altaffilmark{1}, STEPHEN  P. REYNOLDS\altaffilmark{3}, JOHN C. RAYMOND\altaffilmark{1}, KAZIMIERZ J. BORKOWSKI\altaffilmark{3}}
\altaffiltext{1}{Harvard-Smithsonian
Center for Astrophysics, CfA} 
\altaffiltext{2}{Department of Astronomy, School of Physics and Astronomy, University of Minnesota}
\altaffiltext{3}{Department of Physics, North Carolina State University}

\begin{abstract}

G54.1+0.3 is a young pulsar wind nebula (PWN), closely resembling the Crab, for which no thermal shell emission has been detected in X-rays. Recent \textit{Spitzer} observations revealed an infrared (IR) shell containing a dozen point sources arranged in a ring-like structure, previously proposed to be young stellar objects. An extended knot of emission located in the NW part of the shell appears to be aligned with the pulsar's X-ray jet, suggesting a possible interaction with the shell material. Surprisingly, the IRS spectrum of the knot resembles the spectrum of freshly formed dust in Cas A, and is dominated by an unidentified dust emission feature at 21 $\micron$. The spectra of the shell also contain various emission lines and show that some are significantly broadened, suggesting that they originate in rapidly expanding supernova (SN) ejecta. We present the first evidence that the PWN is driving shocks into expanding SN ejecta and we propose an alternative explanation for the origin of the IR emission in which the shell is composed entirely of SN ejecta. In this scenario, the freshly formed SN dust is being heated by early-type stars belonging to a cluster in which the SN exploded. Simple dust models show that this interpretation can give rise to the observed shell emission and the IR point sources. 

\end{abstract}


\section{INTRODUCTION} \label{intro}

G54.1+0.3 is a pulsar wind nebula (PWN) with properties very similar to the Crab. Radio observations show a nebula with a flat spectrum, approximately 2\arcmin\ in extent \citep{rei84,gre85,vel88}, and no evidence of the supernova remnant (SNR) shell. While the extent of the  X-ray nebula in the Crab is 3 times smaller than in the radio, non-thermal X-ray emission from G54.1+0.1 extends out to the radio boundary \citep{sew89,lu01}. \textit{Chandra} observations of G54.1+0.3 revealed a point source surrounded by a ring and torus, and a nebula elongated in the E/W direction \citep{lu02}. All components have a power-law spectrum with an index steepening with increasing distance from the pulsar. There is no evidence for a thermal component associated with supernova (SN) ejecta or the swept up interstellar medium (ISM). 

The 137 ms pulsar J1930+1852 that powers the nebula was discovered by \citet{cam02}, who calculate a characteristic age of 2900 yr. Considering a possible range of braking indices and initial spin periods, \citet{cam02} estimate the age of G54.1+0.3 to be between 1500 and 6000 yr. From the equations describing  PWN evolution, \citet{che05} calculates an age of 1500 yr and and an initial spin period $P_0$ of 100 ms. Based on HI line emission and absorption measurements, the distance to G54.1+0.3 is in the 5--9 kpc range \citep{wei08,lea08}, while the pulsar's dispersion measure implies a distance less than or equal to 8 kpc \citep{cam02}. \citet{lea08} recently suggest a morphological association between the nebula and a CO molecular cloud at a distance of 6.2$\pm$0.1 kpc. However, the absence of X-ray thermal emission and the lack of evidence for an interactin of the SNR with the cloud argue against this association. In this paper, we assume a distance of 6 kpc for G54.1+0.3, at which $\rm 10\arcsec \approx 0.3 \: pc$.

While there has been no previous evidence for an interaction of the PWN with SN ejecta, circumstellar material (CSM), or the ISM in G54.1+0.3, a recent \textit{Spitzer} discovery of an infrared (IR) shell surrounding the PWN may be the first evidence of such an interaction \citep{sla08,koo08}. The first IR detection from G54.1+0.3 was made with the Infrared Astronomical Satellite (IRAS) at 25 \micron, 60 \micron, and 100 \micron, and the emission was consistent with a cold dust component with a temperature of $\sim$ 30 K \citep{are89,sak92}. \textit{Spitzer} imaging revealed an IR shell surrounding the PWN in which a dozen point sources are arranged in a ring-like structure, suggested to be young stellar objects (YSOs) whose formation was triggered in the late stages of the progenitor's life \citep{koo08}. The morphological association between G54.1+0.3 and the shell is evident in the IR images and suggests that the PWN may be interacting with the shell material. The nature of the shell still remains unclear. While the morphology suggests that the shell is SN ejecta and freshly formed dust swept up by the PWN, presence of YSOs would require that it be a preexisting circumstellar shell or a molecular cloud. In this paper, we present deep \textit{Chandra} imaging of G54.1+0.3 and \textit{Spitzer} IR imaging and spectroscopy of the associated shell that provide new evidence that the shell emission originates from SN ejecta and dust. 

The paper is organized as follows; Section \ref{obsv} describes the
observations and reduction of the \textit{Chandra} and
\textit{Spitzer} data, Section \ref{analysis} describes the basic
analysis and fitting of the X-ray spectra and IR line emission,
Section \ref{lineemission} presents a detailed analysis of the IR
spectroscopy, Section \ref{dustemission} discusses the dust emission, Section \ref{interp} describes a new interpretation for the origin of the IR shell and the embedded point sources, and finally, Section \ref{concl} summarizes the conclusions of our study.

\section{OBSERVATIONS AND DATA REDUCTION} \label{obsv}

\subsection{Chandra}

Observations of G54.1+0.3 were taken with the Advanced CCD Imaging Spectrometer, ACIS-S, on board the \chandra X-ray observatory on 2008, July 8, 10, 12, and 15, under the observation ID numbers 9886, 9108, 9109, and 9887. The observations were carried out using the FAINT mode and the corresponding exposure times were 66.17, 35.11, 164.33, and 25.16 ks, for a total exposure time of 290.77 ks. The standard data reduction and cleaning were performed using Ciao Version 3.4. 

\begin{figure}
\epsscale{1.2} \plotone{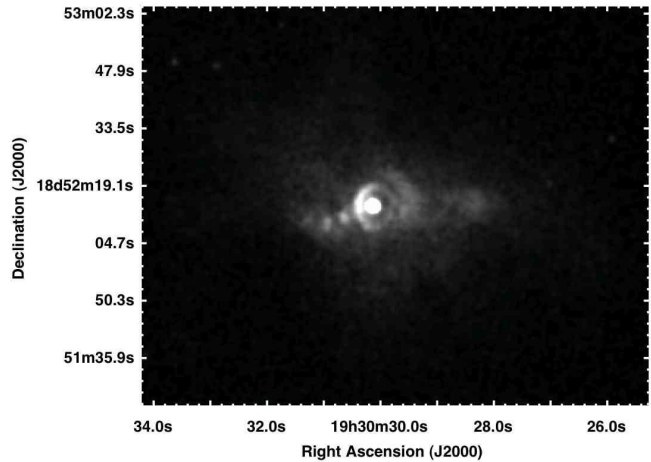} \caption{\label{chandra}Chandra ACIS
image of G54.1+0.3 in the 0.3--10.0 keV band, shown on a linear scale
(0 to 120 counts).}
\end{figure}

\begin{figure}
\epsscale{1.2} \plotone{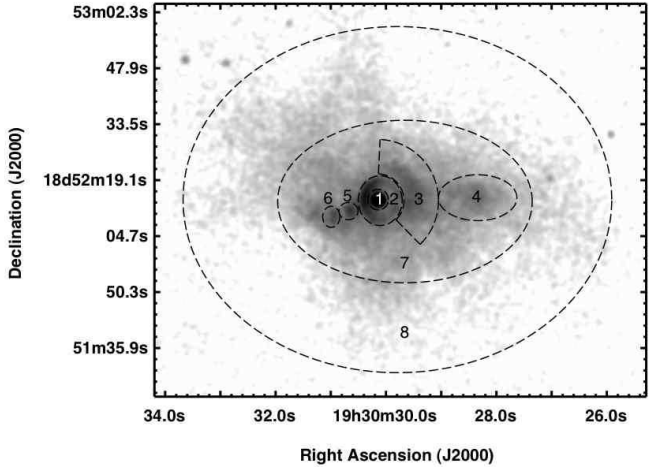} \caption{\label{chandraaps}Extraction
regions for the X-ray spectra, overlaid on the smoothed Chandra ACIS image of
G54.1+0.3 in the 0.3--10.0 keV band (shown on a 0--230 logarithmic scale).The fitted parameters for each of the numbered regions are summarized in Table \ref{chandratab}.}
\end{figure}

The cleaned \chandra datasets were merged into a single event file
using the \textit{merge\_all} task in \textit{Ciao} version 3.4. The
final merged image is shown in Figure \ref{chandra}. Spectra were
extracted from the merged file from eight different regions covering
the PWN and one background region. The extraction regions are similar
to those in \citet{lu02} for the purpose of comparison, and are shown
in Figure \ref{chandraaps}. Because the observations were carried out
close in time and have the same chip orientations and positions on the
sky, the corresponding effective area and spectral response files were
generated from a single  observation with ID 9109.  Source spectra
were background subtracted and grouped to include a minimum of 75
counts in each bin for region 1, 50 counts in each bin for regions 2,
3, 4, 7, and 8, and 20 counts in each bin for regions 5 and 6. The
observations are summarized in Table \ref{data}.

\begin{deluxetable}{lrlc}
\tablecolumns{4} \tablewidth{0pc} \tablecaption{\label{data}CHANDRA AND SPITZER OBSERVATIONS}
\tablehead{
 \colhead{Detector/} & \colhead{Prog.} & \colhead{Observation} & \colhead{Exposure} \\
\colhead{Channel} & \colhead{ID} & \colhead{Date} & \colhead{Time}
}
\startdata

\cutinhead{\textit{Chandra}}
ACIS-S & 9886 & 2008 Jul 8   & 66.17 ks \\
              & 9108 & 2008 Jul 10 & 35.11 ks \\
              & 9109 & 2008 Jul 12 & 164.33 ks \\
              & 9887 & 2008 Jul 15 & 25.16 ks \\
\cutinhead{\textit{Spitzer}}
IRAC & 3647 & 2005 Oct 21 & 2 frames $\times$ 30 s \\
MIPS 24 & 3647 & 2005 May 15 & 20 cycles $\times$ 30 s \\
IRS SL & 40736 & 2007 Nov 05 & 5 cycles $\times$ 14 s \\
IRS LL & 40736 & 2007 Nov  05 & 5 cycles $\times$ 6 s \\
IRS SH & 40736 & 2007 Nov 05 & 10 cycles $\times$ 120 s \\
IRS LH & 40736 & 2007 Nov 05 & 5 cycles $\times$ 6 s \\
\enddata
\tablecomments{The MIPS 70 \micron\ image was obtained from the MIPSGAL survey \citep{car09}.}
\end{deluxetable}

\subsection{Spitzer}\label{spitzer}

Infrared spectroscopy of the IR shell was carried out with all modules of the Infrared Spectrograph (IRS) \citep{hou04} aboard \spitzer on 2007, Nov. 05, covering the 5-37 $\micron$ wavelength range (program ID 40736). The location of the shell with respect to the PWN is shown in Figure \ref{3color}, and the positions of the IRS slits are shown in Figure \ref{slits}. The low resolution spectra were taken at position 19h30m28.0s, +18$^{\circ}$52\arcmin15.60\arcsec, centered on the diffuse shell emission, but also covering the bright knot in the northwest with the long-low (LL) module. The low resolution observations had 5 cycles for each module, with an exposure time of 14 s and 6 s each for the short-low (SL) and LL modules, respectively. The high resolution spectra were taken at two different positions; the diffuse shell emission  at 19h30m28.0s, +18$^{\circ}$52\arcmin15.60\arcsec, and the bright knot at 19h30m26.40s, +18$^{\circ}$52\arcmin07.00\arcsec. The locations of the slits for the background spectrum are also shown in Fig \ref{slits}. The short-high (SH) spectra were obtained using 10 cycles of 120 s, and the long-high (LH) using 5 cycles of 6 s each.

Data were processed with pipeline version S17.0.4 and cleaned using IRS Rogue Pixel Mask Editing and Image Cleaning software (IRSCLEAN1.9). The background was subtracted from the high resolution data, individual Basic Calibrated Data (BCDs) of each nod were median-combined, and spectra were extracted with the Spitzer IRS Custom Extractor (SPICE) using full slit extractions with the extended source calibration for both low and high resolution modules. The subsequent analysis, including averaging of the individual nods, background subtraction for the low resolution modules, trimming of the order edges, and fitting of the emission lines, was carried out using the Spectroscopic Modeling, Analysis and Reduction Tool (SMART) \citep{hig04}. For the low resolution modules, the spectra extracted from the off-source nods were used for the background. 
We have also included the imaging data from the Infrared Array Camera
(IRAC) and Multiband Imaging Photometer (MIPS) aboard
\textit{Spitzer}, shown in Figures \ref{3color} and \ref{irac}. The
IRAC observations at 3.6, 4.5, 5.8, and 8.0 $\micron$ were carried out
on 2005, Oct. 21, under the program ID 3647, using two 30 s
frames. The data were processed with the pipeline version S14.0.0 and
reduced with the MOsaicer and Point source EXtractor (MOPEX) version
18.1.5. The MIPS 24 $\micron$ observation was carried out on 2005, May
15, under the program ID 3647, using 20 cycles of 30 s, and was
processed with the pipeline version S16.1.0 \citep{sla08}. The MIPS 70
$\micron$ image of G54.1+0.3 was obtained from the MIPSGAL survey
\citep{car09}. \textit{Spitzer} observations are summarized in Table \ref{data}.

\begin{deluxetable*}{clcccccc}
\tablecolumns{7} \tablewidth{0pc} \tablecaption{\label{chandratab}CHANDRA ACIS SPECTRAL FITTING RESULTS}
\tablehead{
\colhead{Region} & \colhead{Description} & \colhead{Area(arcsec$^2$)} & \colhead{Photon Index} & \colhead{$N_H$($10^{22}cm^{-2}$)} & \colhead{$F_X$(observed)} & \colhead{$F_X$(unabsorbed)} & \colhead{Reduced $\chi^2$}}
\startdata
1 & Pulsar & 22.03 & 1.44 $\pm$ 0.04  & (1.95) & 2.10E-12 & 3.26E-12 & 1.03 \\
2 & Ring & 90.77& 1.86 $\pm$ 0.05  & 1.95 $\pm$ 0.04 & 5.88E-13 & 1.21E-12 & 0.66 \\
3 & West arc & 230.9 & 1.89 $\pm$ 0.05  & ... & 6.64E-13 & 1.40E-12 & 0.92 \\
4 & West jet & 185.4 & 1.79 $\pm$ 0.06  & ... & 4.06E-13 & 7.88E-13 & 1.08 \\
5 & East knot 1 & 16.94 & 1.90 $\pm$ 0.12  & ... & 8.94E-14 & 1.90E-13 & 0.63 \\
6 & East knot 2 & 19.12 & 1.97 $\pm$ 0.13  & ... & 7.62E-14 & 1.73E-13 & 0.70 \\
7 & Inner nebula & 1586 & 2.05 $\pm$ 0.04  & ... & 1.84E-12 & 4.45E-12 & 0.94 \\
8 & Outer nebula & 5559 & 2.20 $\pm$ 0.04  & ... & 1.18E-12 & 3.34E-12 & 0.83 \\

\enddata
\tablecomments{The extraction regions are shown in Figure \ref{chandraaps}. Listed uncertainties are 1.6 $\sigma$ (90 \% confidence) statistical uncertainties from the fit. The $N_H$ value listed for Region 2 is the result of the simultaneous fit for regions 2--8. The fluxes were calculated in the 0.3-10 keV band and are in the units of $\rm erg\:cm^{-2}\:s^{-1}$.}
\end{deluxetable*}


\section{ANALYSIS}\label{analysis}

\subsection{X-ray}

The deep \chandra ACIS image of G54.1+0.3 in the 0.3-10 keV band is shown in Figure \ref{chandra}. The central pulsar is surrounded by a 5\farcs7 by 3\farcs7 X-ray ring, with the long axis oriented roughly north-south, suggesting an inclination angle of about 40 degrees \citep{lu02}. The pulsar and the ring are embedded in a diffuse nebula, 2\farcm0 by 1\farcm3 in size, which appears to have a bipolar elongation, running roughly east-west, perpendicular to the apparent plane of the ring. 

Spectra were extracted from the \chandra data using the apertures shown in Figure \ref{chandraaps}. The regions are similar to those in \citet{lu02} and include the pulsar in region 1, the surrounding ring in region 2, the western arc in region 3, the western jet in region 4, two bright knots of emission in the east in regions 5 and 6, and fainter extended emission between these structures and in the outer regions of the PWN in regions 7 and 8, respectively. The background spectrum was extracted from an elliptical region with an area of 7710 square arcseconds, centered at the position 19h30m29.6s, +18$^{\circ}$50\arcmin35.25\arcsec.
The spectrum for each of the regions was fitted with an absorbed power
law model using the \textit{Ciao} 3.4 \textit{Sherpa} software. The
fitting of region 1 also included a pileup model, since the point
source count rate of  0.077 $s^{-1}$ is expected to cause significant
pileup that affects the spectral parameters \citep{dav01}. The
absorbing column density was fit simultaneously for regions 2--8, and
the column density for the region encompassing the point source was
then fixed to this value. The fitting results are summarized in Table
\ref{chandratab}, and the spectra are shown in Figures
\ref{xrayspectra1} and \ref{xrayspectra2}. We find an absorption column density of $(1.95\pm0.04)\times10^{22}{\rm\ cm^{-2}}$, somewhat greater than that of \citet{lu02}, who found $N_{H}=(1.6\pm0.1)\times10^{22}{\rm\ cm^{-2}}$. We find a similar value to that of \citet{lu02} if we include region 1 in the joint fit and do not account for pileup. When the spectrum of each region is fit separately, the $N_H(10^{22}{\rm\ cm^{-2}})$ in regions 2--8 ranges from 1.88--2.19, while the best fit for region 1 gives an $N_H(10^{22}{\rm\ cm^{-2}})$ of 1.54. This suggests that a more accurate estimate for $N_H$ is achieved if the piled-up region 1 is excluded from the fit. 

\begin{deluxetable}{lcc}
\tablecolumns{4} \tablewidth{0pc} \tablecaption{\label{iractab}IRAC AND MIPS FLUXES}
\tablehead{
\colhead{Wavelength} & \colhead{Flux} & \colhead{Extinction Corrected} \\
\colhead{($\micron$)} & \colhead{(Jy)} & \colhead{Flux (Jy)}
}
\startdata
5.8 & $\sim$ 0.2 &  $\sim$ 0.3 \\
8.0 & $\sim$ 0.6 &  $\sim$ 1.0 \\
24 & 23.7 $\pm$ 2.4 & 40 $\pm$ 4 \\
24 -- \textit{IR knot} & 3.8 $\pm$ 0.4 & 6.5 $\pm$ 0.7 \\
70 & 76 $\pm$ 15 & \nodata 

\enddata
\tablecomments{IRAC fluxes were estimated based on the average flux density at selected positions scaled to the size of the shell. The uncertainties include IRAC and MIPS calibration uncertainties, but do not account for the uncertainties in the extinction correction. The extinction correction was applied using the extinction curve of \citet{chi06}. }
\end{deluxetable}

The spectral fits show that the outer regions of the diffuse emission in G54.1+0.3 have a softer spectrum than the interior regions. We find no evidence of a thermal component associated with shock-heated ejecta or ISM in the X-ray spectra. The new estimate of the absorbing column density results in a slightly higher photon index for the pulsar, $\Gamma=1.44\pm0.04$. The ring and E/W elongations have similar photon indices, which are harder (smaller) than those in the outer diffuse nebula, suggesting lower synchrotron losses or more particle injection in these regions. There is no evidence for emission lines in any of the spectra. The parameters derived from the initial analysis of the \textit{Chandra} data are used in our analysis of the IR observations. A more detailed study of the X-ray data will be presented in a separate publication.


\begin{figure}
\epsscale{1.2} \plotone{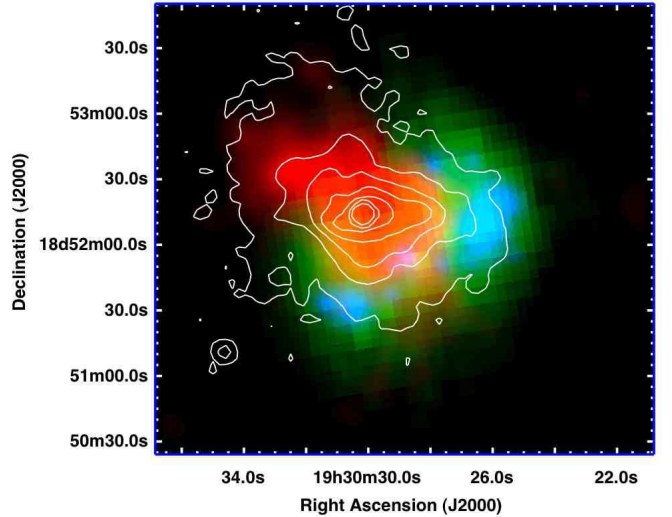} \caption{\label{3color}Three color image of G54.1+0.3, where the MIPS 24 $\micron$ emission is shown in blue, MIPS 70 $\micron$ emission in green, and radio emission in red. The \chandra X-ray contours are shown in white.}
\end{figure}

\subsection{IR Morphology}

The \spitzer IRAC and MIPS images of G54.1+0.3 reveal an infrared shell south and west of the PWN, with the X-ray nebula filling the cavity of the shell. The shell is detected at 5.8, 8.0, 24, and 70 $\micron$, and is approximately 1\farcm5 in radius at 70 \micron. A shell structure is evident in the 5.8 $\micron$ and 8.0 $\micron$ IRAC images, shown in Figures \ref{irac}a and \ref{irac}b, along with significant emission that extends far from the PWN to the north that may be associated with Polycyclic Aromatic Hydrocarbon (PAH) emission unrelated to G54.1+0.3. The structure of the shell in the MIPS images appears to have a a brighter western lobe, and a fainter southeastern lobe (Figure \ref{irac}c and \ref{irac}d). The X-ray and radio emission from the PWN fills the cavity of the shell. The western jet structure, in particular, appears to extend directly into a void region in the IR nebula, terminating at the brightest region in the 24 $\micron$ MIPS image, i.e. the IR knot. This is best shown in Figure \ref{3color}, where the MIPS 24 $\micron$ emission is shown in blue, MIPS 70 $\micron$ emission in green, radio emission in red, and \chandra X-ray contours in white. The most interesting features of the 24 $\micron$ image are 11 point sources embedded in the diffuse shell that are arranged in a ring-like structure \citep{koo08}. The positions of the point sources are marked in Figure \ref{slits}. The two northern point sources are outside of the ring structure, but appear to be located at the tips of faint ridges of emission that trace back to the shell structure. The IRAC and MIPS fluxes from the IR shell and the knot are listed in Table \ref{iractab}. Since the shell region in the IRAC images is crowded with stars, the fluxes were estimated by scaling the average flux density from several positions on the shell by the total area of 4 square arcminutes. In measuring the flux from the IR knot, we used a background region from the diffuse shell emission east of the knot.

\subsection{IR Spectroscopy}

\begin{figure}
\epsscale{1.15} \plotone{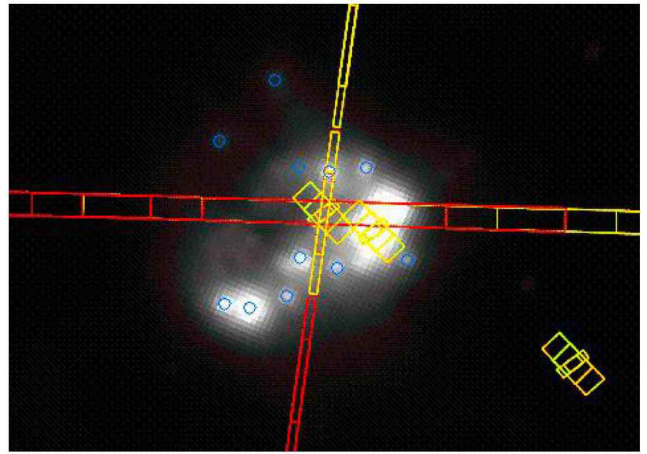} \caption{\label{slits}Positions of the IRS slits overlaid on the MIPS 24 \micron\ image. The LL slits are the wide slits oriented in the East/West direction. The SL slits are oriented North/South. The LH and SH slits are the smaller slits in yellow, positioned at three different locations; the diffuse shell emission at the same location as the low resolutions slits( \textit{position 1}), the extended IR knot in the west (\textit{position 2}), and the background region located off of the shell to the west. The point sources embedded in the diffuse shell that are evident in the MIPS 24 \micron\ image are marked by blue circles.}
\end{figure}

\begin{figure*}
\epsscale{1.0} \plotone{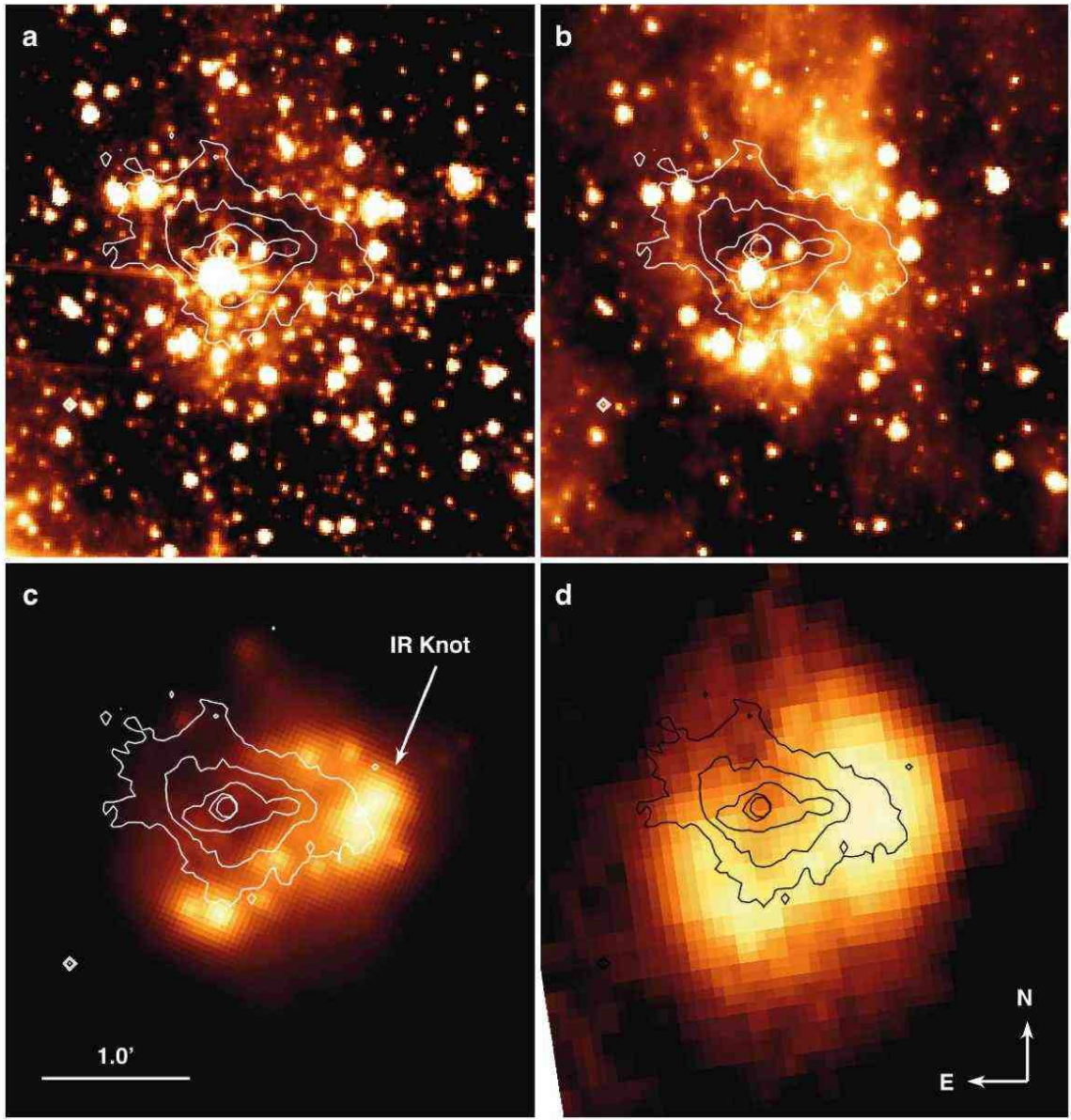} \caption{\label{irac}Spitzer imaging of the shell surrounding the PWN in G54.1+0.3. The shell is detected in the IRAC 5.8 \micron\ (panel a) and 8.0 \micron\ (panel b) bands. The MIPS 24 \micron\ and 70 \micron\ images are shown in panels c and d, respectively. The IR knot is indicated by the white arrow. The overlaid contours represent the Chandra ACIS contour of the PWN.}
\end{figure*}

\begin{figure*}
\epsscale{1.1} \plotone{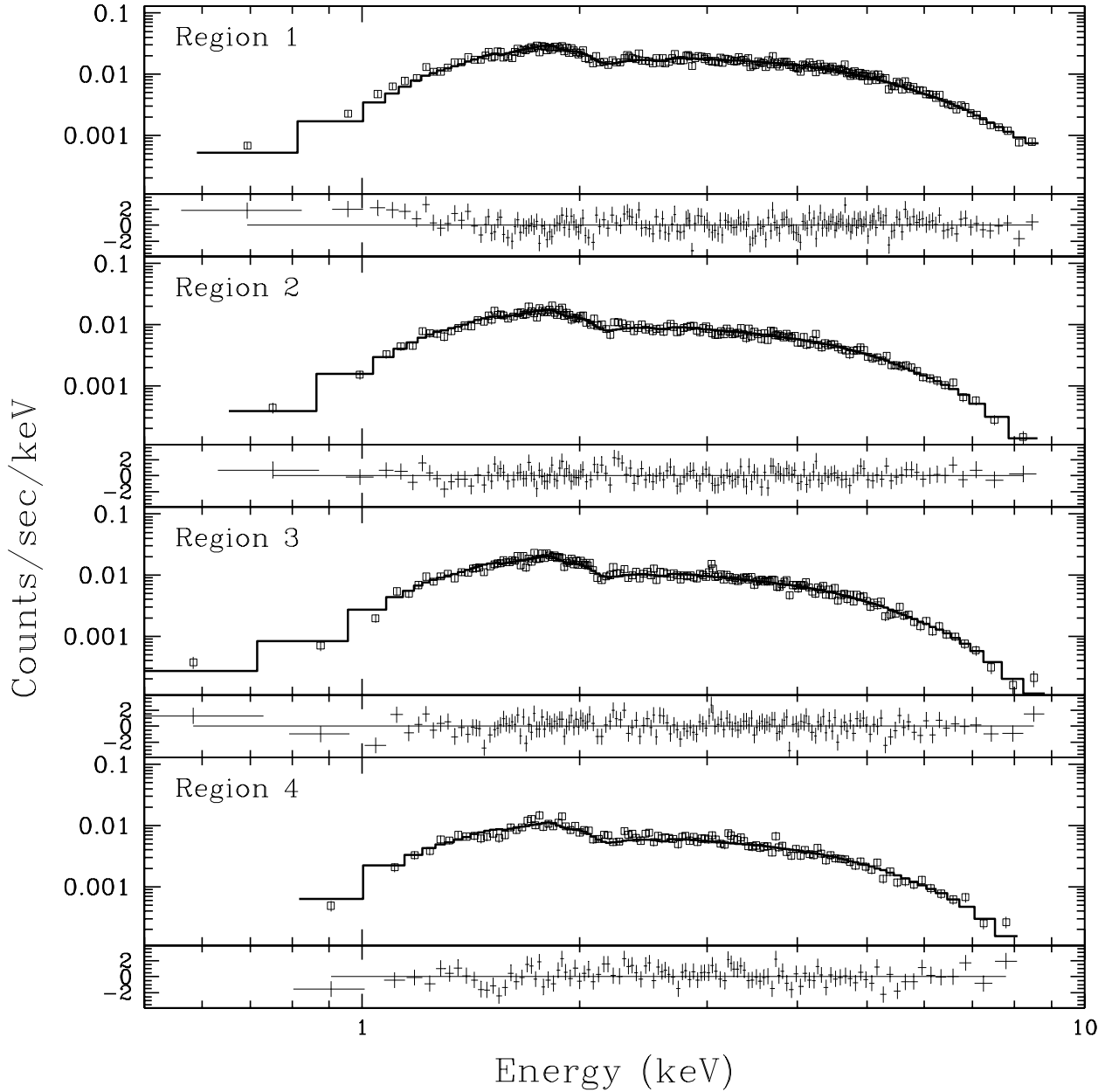}
\caption{\label{xrayspectra1}\textit{Chandra} X-ray spectra of the PWN
in G54.1+0.3 extracted from regions 1--4, shown in Figure \ref{chandraaps}. The spectra in all regions are well fit by an absorbed power-law model with the parameters listed in Table \ref{chandratab}.}
\end{figure*}

\begin{figure*}
\epsscale{1.1}
\plotone{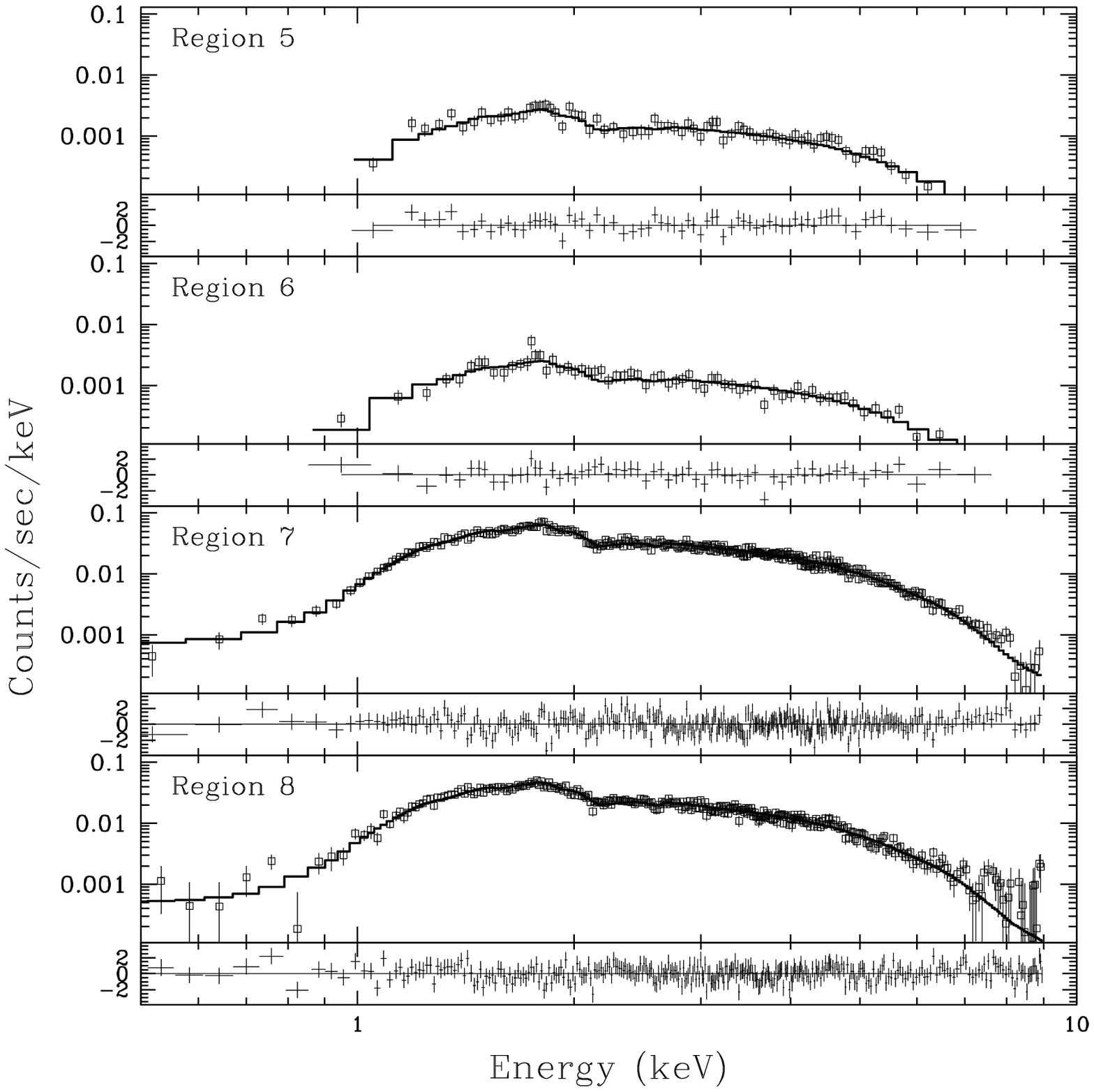}\caption{\label{xrayspectra2}\textit{Chandra} X-ray spectra of the PWN in G54.1+0.3 extracted from regions 5--8, shown in Figure \ref{chandraaps}. The spectra in all regions are well fit by an absorbed power-law model with the parameters listed in Table \ref{chandratab}.}
\end{figure*}

The high and low resolution IR spectra of G54.1+0.3 are shown in Figures \ref{highres} and \ref{lowres}, and the positions of the IRS slits are shown in Figure \ref{slits}. The low resolution spectrum shows a rising continuum longward of 15 \micron, and a broad emission feature peaking at approximately 21 \micron. This feature is more pronounced in the high resolution spectrum of the bright knot, while it is less pronounced in the spectrum from the diffuse shell (Figure \ref{highres}). The high resolution spectra also reveal a broad emission feature around 12 \micron, and a narrow emission feature around 11.3 \micron\ that may be associated with PAH emission or other carbonaceous grains. The high resolution slit from which the background spectrum was extracted is located in a region of low background emission, and it underestimates the contribution from the emission in the vicinity of the IR shell that is evident in the IRAC 8 \micron\ image (Figure \ref{irac}b). However, the background regions for the low resolution spectrum are more representative of the local background emission. Since the low resolution spectra were extracted from the entire SL and LL slits, they are not characteristic of the point-to-point spectrum along the slits that varies spatially. The sharp peak around 7.5 \micron\ is an artifact caused by a mismatch between orders, and was trimmed from the residual source spectrum.


\begin{deluxetable*}{lccccccc}
\tablecolumns{8} \tablewidth{0pc} \tablecaption{\label{spitzertab}IRS LINE FITS}
\tablehead{
\colhead{Line ID} & \colhead{Line Center} & \colhead{Line Flux} & \colhead{De-reddened Flux} & \colhead{FWHM} & \colhead{FWHM} & True FWHM & \colhead{Shift} \\
\colhead{} & \colhead{(\micron)} & \colhead{($10^{-6} erg/cm^{2}/s/sr$)} & \colhead{($10^{-6} erg/cm^{2}/s/sr$)} & \colhead{(\micron)} & \colhead{($km/s$)} & \colhead{($km/s$)} & \colhead{($km/s$)}
}
\startdata
\sidehead{Position 1}
H$_2$ (12.2786) &  12.2775 $\pm$  0.0013 & 1.24 $\pm$ 0.31 & 2.15 $\pm$ 0.54 & 0.018 $\pm$ 0.001 &  435 $\pm$ 24 & ...  & -29 $\pm$ 35 \\
$[$\ion{Ne}{2}$]$ (12.8135) & 12.8138 $\pm$ 0.0008 & 11.15 $\pm$ 0.76 & 18.72 $\pm$ 1.28 & 0.024 $\pm$ 0.001 &  568 $\pm$ 24 & 279 $\pm$ 104 & 5 $\pm$ 20 \\
$[$\ion{Cl}{2}$]$ (14.3678) & 14.3721 $\pm$ 0.0005 & 4.59 $\pm$ 0.16 & 7.79 $\pm$ 0.27 & 0.049 $\pm$ 0.001 & 1029 $\pm$ 21 & 902 $\pm$ 32 & 89 $\pm$ 9 \\
H$_2$ (17.0348) & 17.0385 $\pm$ 0.0004 & 2.74 $\pm$ 0.10 & 5.10 $\pm$ 0.19 & 0.030 $\pm$ 0.001 &  525 $\pm$ 18 & 177 $\pm$ 165  & 65 $\pm$ 6 \\
$[$\ion{S}{3}$]$ (18.7130) & 18.7195 $\pm$ 0.0010 & 13.51 $\pm$ 0.83 & 26.61 $\pm$ 1.64 & 0.051 $\pm$ 0.001 &  823 $\pm$ 16 & 658 $\pm$ 44 & 104 $\pm$ 16 \\
$[$\ion{Fe}{2}$]$ (25.9883) & 26.0048 $\pm$ 0.0084 & 4.62 $\pm$ 1.26 & 7.33 $\pm$ 1.99 & 0.095 $\pm$ 0.008 & 1089 $\pm$ 92 & 966 $\pm$ 33 & 191 $\pm$ 97 \\
$[$\ion{S}{3}$]$ (33.4810) & 33.4890 $\pm$ 0.0026 & 25.28 $\pm$ 2.67 & 35.94 $\pm$ 3.79 & 0.075 $\pm$ 0.003 & 675 $\pm$ 27 & 450 $\pm$ 70 & 72 $\pm$ 23 \\
$[$\ion{Si}{2}$]$ (34.8152) & 34.8174 $\pm$ 0.0040 & 72.69 $\pm$ 6.67 & 102.76 $\pm$ 9.43 & 0.136 $\pm$ 0.004 & 1172 $\pm$ 34 & 1059 $\pm$ 30 & 19 $\pm$ 35 \\
\sidehead{Position 2}
$[$\ion{S}{4}$]$ (10.5105) & 10.5164 $\pm$ 0.0019 & 1.07 $\pm$  0.19  & 2.47 $\pm$ 0.43 & 0.021 $\pm$ 0.001 &  596 $\pm$ 28  & 334 $\pm$ 87 & 168 $\pm$ 55 \\
$[$\ion{Ne}{2}$]$ (12.8135) & 12.8123 $\pm$ 0.0003 & 9.74 $\pm$ 0.30 & 16.36 $\pm$ 0.50 & 0.029 $\pm$ 0.001 &  675 $\pm$ 23 & 450 $\pm$ 70 & -28 $\pm$ 6 \\
$[$\ion{Cl}{2}$]$ (14.3678) & 14.3689 $\pm$ 0.0011 & 3.12 $\pm$ 0.28 & 5.28 $\pm$ 0.48 & 0.037 $\pm$ 0.001 & 780 $\pm$ 21 & 604 $\pm$ 48 & 23 $\pm$ 21 \\
H$_2$ (17.0348) & 17.0392 $\pm$ 0.0016 & 1.18 $\pm$ 0.22 & 2.20 $\pm$ 0.40 & 0.028 $\pm$ 0.002 &  490 $\pm$ 35 & ... & 77 $\pm$ 29 \\
$[$\ion{S}{3}$]$ (18.7130) & 18.7212 $\pm$ 0.0006 & 33.92 $\pm$ 1.49 & 66.81 $\pm$ 2.94 & 0.041 $\pm$ 0.001 &  651 $\pm$ 16 & 424 $\pm$ 69 & 132 $\pm$ 9 \\
$[$\ion{S}{3}$]$ (33.4810) & 33.4939 $\pm$ 0.0011 & 48.08 $\pm$ 2.68 & 68.36 $\pm$ 3.80 & 0.059 $\pm$ 0.001 & 528 $\pm$ 9 & 160 $\pm$ 198 & 116 $\pm$ 10 \\
$[$\ion{Si}{2}$]$ (34.8152) & 34.8229 $\pm$ 0.0046 & 57.30 $\pm$ 7.56 & 81.00 $\pm$ 10.68 & 0.108 $\pm$ 0.005 & 934 $\pm$ 43 & 786 $\pm$ 40 & 66 $\pm$ 40 \\


\enddata
\tablecomments{ Listed uncertainties are 1-$\sigma$ statistical uncertainties from the fit and do not include the IRS wavelength and flux calibration uncertainties.}
\end{deluxetable*}

There are a total of ten emission lines present in the spectrum of the IR shell; the  molecular hydrogen H$_2$ S(2) (12.3 \micron), and H$_2$ S(1) (17.0 \micron) lines, and ionic lines of [\ion{Ar}{2}] (6.99 \micron), [\ion{S}{4}] (10.51 \micron), [\ion{Ne}{2}] (12.81 \micron), [\ion{Cl}{2}] (14.37 \micron), [\ion{S}{3}] (18.71 \micron), [\ion{Fe}{2}] (25.99 \micron),  [\ion{S}{3}] (33.48 \micron), and  [\ion{Si}{2}] (34.82 \micron). The position of the line at 14.37 \micron\ that we identify as [\ion{Cl}{2}] is near the position of [\ion{Ne}{5}] 14.32 \micron\, but since we do not detect \ion{Ne}{3} in our spectrum, we are confident that the line is produced by chlorine.
The line-fitting parameters for the LH and SH modules are listed in Table \ref{spitzertab}, where positions 1 and 2 correspond to spectra from the diffuse shell and the bright IR knot, respectively (see Figure \ref{slits}). Line-fitting parameters from the overlapping region of SL and LL modules are listed in Table \ref{lowspectab}. The extinction correction was applied using the IR extinction curve of \citet{chi06}, and the relation $N_{H}/A_{K}=1.821\times10^{22}{\rm\ cm^{-2}}$ \citep{dra89}. As shown in Figure \ref{lowres}, background emission significantly contributes to the observed lines of Ne, S and Si, implying that the uncertainties on their measured line intensities are probably larger than the statistical uncertainties quoted in Table \ref{spitzertab}. A number of emission lines in the high resolution spectrum show evidence of broadening with respect to the spectral resolution of IRS of $\lambda/{\Delta\lambda}\sim600$, including Cl, S, Fe, and Si. The broadened line profiles of the Si and S lines are shown in Figure \ref{profiles}.

\begin{deluxetable}{lccc}
\tablecolumns{4} \tablewidth{0pc} \tablecaption{\label{lowspectab}IRS LINE FITS IN LL \& SL OVERLAP REGION}
\tablehead{
\colhead{Line ID} & \colhead{Line Center ($\micron$)} & \colhead{Line Flux} & \colhead{De-reddened}
}
\startdata
\sidehead{Region 1}
$[$\ion{Ar}{2}$]$ (6.9853) & 6.993 $\pm$ 0.001& 84.9 $\pm$ 4.9 & 122.4 $\pm$ 7.1 \\
$[$\ion{Ne}{2}$]$ (12.8135) & 12.813 $\pm$ 0.013 & 6.4 $\pm$ 1.9 & 10.8 $\pm$ 3.3 \\
$[$\ion{Cl}{2}$]$  (14.3678) & 14.401 $\pm$ 0.013 & 3.2 $\pm$ 1.3 & 5.4 $\pm$ 2.2 \\
$[$\ion{S}{3}$]$ (18.7130) & 18.733 $\pm$ 0.004 & 11.9 $\pm$ 1.1 & 23.5 $\pm$ 2.2 \\
$[$\ion{S}{3}$]$ (33.4810) & 33.539 $\pm$ 0.029 & 33.7 $\pm$ 8.4 & 47.9 $\pm$ 11.9  \\
$[$\ion{Si}{2}$]$ (34.8152) & 34.889 $\pm$ 0.015 & 79.3 $\pm$ 10.6 & 112.1 $\pm$ 14.9  \\

\enddata
\tablecomments{Fluxes are in units of $\rm 10^{-6} erg/cm^{2}/s/sr$. Listed uncertainties are 1-$\sigma$ statistical uncertainties from the fit and do not include the IRS wavelength and flux calibration uncertainties.}
\end{deluxetable}

For the purpose of examining the spatial variation in the emission line intensities across the LL slits, we created 2-dimensional spectral line maps using the CUbe Builder for IRS Spectral Maps (CUBISM) software \citep{smi07}. The maps were made by selecting a desired wavelength region for the source flux and subtracting a wavelength-weighted background, averaged across a specified wavelength region. The maps were created by selecting a source flux from 20.9--21.9 $\micron$, with a continuum  from 20.5--20.7 $\micron$ and 22.3--22.6 $\micron$ to show the spatial variation of the 21 $\micron$ feature; a source flux between 18.4--18.9 \micron\, with a continuum between 18.0--18.4 \micron\ and 18.9--19.0 \micron\ for the [\ion{S}{3}] line at 18.7 \micron; a source flux between 33.1--33.9 $\micron$, with a continuum between 31.9--32.9 $\micron$ and 33.9--34.2 $\micron$ for the [\ion{S}{3}] line at 33.5 $\micron$; and a source flux from 34.6--35.2 $\micron$, with a continuum from 34.1--34.4 $\micron$ for the [\ion{Si}{2}] line at 34.8 $\micron$. The narrow wavelength ranges for the backgrounds were chosen to avoid other emission features and large gradients in the continuum slope. We smoothed the spectral line maps by a 2-pixel Gaussian before plotting the intensity profiles as a function of position along the slit (Figure \ref{spatial}). Spectra were also extracted from 7 spatial positions along the LL slit in 3$\times$2 pixel sub-slits. Due to the low extraction area, only the brightest lines of $S$ and $Si$ were measurable in these spectra. Their intensities are listed in Table \ref{llspectab}, where the position along the LL slit is denoted by the pixel number on the x-axis in Figure \ref{spatial}.                                      


\section{LINE EMISSION} \label{lineemission}

Spectra of the diffuse shell were obtained with all four modules of the IRS (Figure \ref{slits}). The high resolution SH and LH spectra were taken at two different positions on the shell; at the position of diffuse emission near the interface between the shell and the PWN  (position 1), and at the position of the bright IR knot in the northwest (position 2). The LL slit was oriented in the East/West direction and it spans the entire length of the shell, including the IR knot. The SL slits were roughly perpendicular to the LL slits, and span the length of the shell in the North/South direction. 

\subsection{Line Profiles and Broadening}

The high-resolution spectra at both positions in the IR shell show evidence of spectral line broadening. The fitted line parameters for the two positions are summarized in Table \ref{spitzertab}. The expected resolution of LH and SH is $\lambda/{\Delta\lambda}\sim600$, equal to a full width at half maximum (FWHM) of 500 ${\rm\ km\:s^{-1}}$. The [\ion{Cl}{2}], [\ion{Fe}{2}], and [\ion{Si}{2}] lines at position 1 show the largest broadening of over 1000 ${\rm\ km\:s^{-1}}$. The [\ion{S}{3}] lines at this position are broadened to approximately 700--800 ${\rm\ km\:s^{-1}}$. The [\ion{Ne}{2}] line at position 2 shows evidence for slight broadening, while the widths of the  H$_2$ lines are approximately equal to the resolution of the IRS. It is likely that inadequate subtraction of the local background from the LH and SH spectra contributes significantly to spectral lines with narrow widths. The true FWHM in Table \ref{spitzertab} was calculated by subtracting the instrumental FWHM of the IRS from the observed FWHM in quadrature \citep[see][]{das08}.


The lines in the spectrum at the position of the IR knot (position 2) are broadened by a lesser amount than at position 1. The [\ion{Cl}{2}] and [\ion{Si}{2}] lines again show the greatest broadening of $\sim$ 800--900 ${\rm\ km\:s^{-1}}$, while the [SIII] lines are only slightly broadened to an average of $\sim$ 600 ${\rm\ km\:s^{-1}}$. We also note that the best-fit centroid of the [\ion{S}{3}] lines at both positions appears to be redshifted by an average of  $\sim$ 120 ${\rm\ km\:s^{-1}}$. Although this shift is not very large considering that the uncertainty of the IRS wavelength calibration for the high-resolution modules is on the order of 100 ${\rm\ km\:s^{-1}}$, it is consistently larger than the shift of all the other observed lines. The exception is the [\ion{Fe}{2}] line with a redshift of 191 $\pm$ 97 ${\rm\ km\:s^{-1}}$, but this line is faint and the shift has a high uncertainty. The profiles of the lines with the highest broadening show evidence for structure. The high resolution line profiles of [\ion{Si}{2}] 34.8 \micron\ and [\ion{S}{3}] 33.5 \micron\ at both positions are shown in Figure \ref{profiles} for comparison. At position 1, the [\ion{Si}{2}] line appears to have a brighter component that peaks at approximately the same wavelength as [\ion{S}{3}], in addition to a fainter blue shifted wing.

\begin{deluxetable*}{lccccc}
\tablecolumns{6} \tablewidth{0pc} \tablecaption{\label{llspectab}IRS LINE FITS ACROSS LL SLIT}
\tablehead{
\colhead{Line ID} & \colhead{Line Center} & \colhead{Line Flux} & \colhead{De-reddened} & \colhead{SIII Line Ratio} & \colhead{Electron Density} \\
\colhead{} & \colhead{(\micron)} & \colhead{($10^{-6} erg/cm^{2}/s/sr$)} & \colhead{($10^{-6} erg/cm^{2}/s/sr$)} & \colhead{(18.7\micron$/$33.5\micron)} & \colhead{($cm^{-3}$)}
}
\startdata
\sidehead{Region 1 (Pixel 14)}
$[$\ion{Si}{2}$]$ (34.8152) & 35.01 $\pm$ 0.02 & 35.4 $\pm$ 4.5 & 49.9 $\pm$ 6.4 & & \\
\sidehead{Region 2 (Pixel 17)}
$[$\ion{S}{3}$]$ (18.7130) & 18.76 $\pm$ 0.01 & 4.7 $\pm$ 0.8 & 9.3 $\pm$ 1.5 & & \\
$[$\ion{S}{3}$]$ (33.4810) & 33.61 $\pm$ 0.02 & 9.5 $\pm$ 1.6 & 13.5 $\pm$ 2.3  & 0.69 $\pm$ 0.16 & $<$ 800 \\
$[$\ion{Si}{2}$]$ (34.8152) & 34.95 $\pm$ 0.02 & 47.6 $\pm$ 7.2 & 67.7 $\pm$ 10.1 & & \\
\sidehead{Region 3 (Pixel 20)}
$[$\ion{S}{3}$]$ (18.7130) & 18.73 $\pm$ 0.01 & 9.1 $\pm$ 2.1 & 18.0 $\pm$ 4.3  &  & \\
$[$\ion{S}{3}$]$ (33.4810) & 33.57 $\pm$ 0.02 & 15.4 $\pm$ 3.8 & 21.8 $\pm$ 5.4  & 0.83 $\pm$ 0.28 & $<$ 1300\\
$[$\ion{Si}{2}$]$ (34.8152) & 34.90 $\pm$ 0.01 & 106.3 $\pm$ 8.4 & 149.9 $\pm$ 11.9 & & \\
\sidehead{Region 4 (Pixel 23)}
$[$\ion{S}{3}$]$ (18.7130) & 18.73 $\pm$ 0.01 & 12.1 $\pm$  1.0 & 23.8 $\pm$ 2.0  & & \\
$[$\ion{S}{3}$]$ (33.4810) & 33.52 $\pm$ 0.02 & 28.9 $\pm$ 6.7 & 41.0 $\pm$ 9.5  & 0.58 $\pm$ 0.14 & $<$ 600 \\
$[$\ion{Si}{2}$]$ (34.8152) & 34.89 $\pm$ 0.01 & 78.2 $\pm$ 10.7 & 110.3 $\pm$ 15.1 &  & \\
\sidehead{Region 5 (Pixel 26)}
$[$\ion{S}{3}$]$ (18.7130) & 18.73 $\pm$ 0.01 & 26.5 $\pm$ 2.0 & 52.1 $\pm$ 4.0 & & \\
$[$\ion{S}{3}$]$ (33.4810) & 33.44 $\pm$ 0.02 & 29.6 $\pm$ 8.1 & 42.0 $\pm$ 11.5 & 1.24 $\pm$ 0.35 & 500 -- 2400 \\
$[$\ion{Si}{2}$]$ (34.8152) & 34.89 $\pm$ 0.02 & 64.6 $\pm$ 8.6 & 91.7 $\pm$ 12.3 & & \\
\sidehead{Region 6 (Pixel 29)}
$[$\ion{S}{3}$]$ (18.7130) & 18.72 $\pm$ 0.01 & 23.9 $\pm$ 3.1 & 47.0 $\pm$ 6.0 & & \\
$[$\ion{S}{3}$]$ (33.4810) & 33.55 $\pm$ 0.01 & 54.9 $\pm$ 7.7 & 77.9 $\pm$ 10.9 & 0.60 $\pm$ 0.11 & $<$ 600 \\
$[$\ion{Si}{2}$]$ (34.8152) & 34.84 $\pm$ 0.01 & 50.5 $\pm$ 7.4 & 71.6 $\pm$ 10.6 & &  \\
\sidehead{Region 7 (Pixel 32)}
$[$\ion{S}{3}$]$ (18.7130) & 18.71 $\pm$ 0.01 & 11.1 $\pm$ 3.7 & 21.8 $\pm$ 7.3 & & \\
$[$\ion{S}{3}$]$ (33.4810) & 33.53 $\pm$ 0.02 & 21.1 $\pm$ 4.2 & 30.0 $\pm$ 6.0 &  0.73 $\pm$ 0.28 & $<$ 1100 \\
$[$\ion{Si}{2}$]$ (34.8152) & 34.90 $\pm$ 0.03 & 48.5 $\pm$ 7.8 & 68.8 $\pm$ 11.1 & & \\

\enddata
\tablecomments{Listed uncertainties are 1-$\sigma$ statistical uncertainties from the fit and do not include the IRS wavelength and flux calibration uncertainties. The ranges of estimated densities assume a temperature range of 100-10000 K, and take into account the statistical uncertainties on the [\ion{S}{3}] line ratios.}
\end{deluxetable*}


\subsection{Spatial Variation in Line Intensities}\label{spatialvariation}

The variation in the surface brightness across the LL slit of the [\ion{S}{3}] 18.7 \micron, [\ion{S}{3}] 33.5 \micron, and the [\ion{Si}{2}] 34.8 \micron\ lines, and the SL spectral line map of the [\ion{Ar}{2}] 6.99 \micron\ line are shown in Figure \ref{spatial}. The peak surface brightness for each line has been normalized to unity, so their relative fluxes are not represented by the plot. The intensities as a function of position for all lines in Figure \ref{spatial} show structure that correlates with the MIPS 24 \micron\ image of the shell. The intensity of the [\ion{S}{3}] 33.5 \micron\ line is well correlated with the 24 \micron\ emission and it appears to stay roughly constant across the western part of the IR shell, with a possible slight rise at the location of the bright IR knot. The spatial intensity profile of the [\ion{S}{3}] 18.7 \micron\ line is similar to the profile of the 21 \micron\ feature and shows a sharp peak at the position of the IR knot, implying an increase in density at this position (see Section \ref{density}).


\begin{figure*}
\epsscale{1.0} \plotone{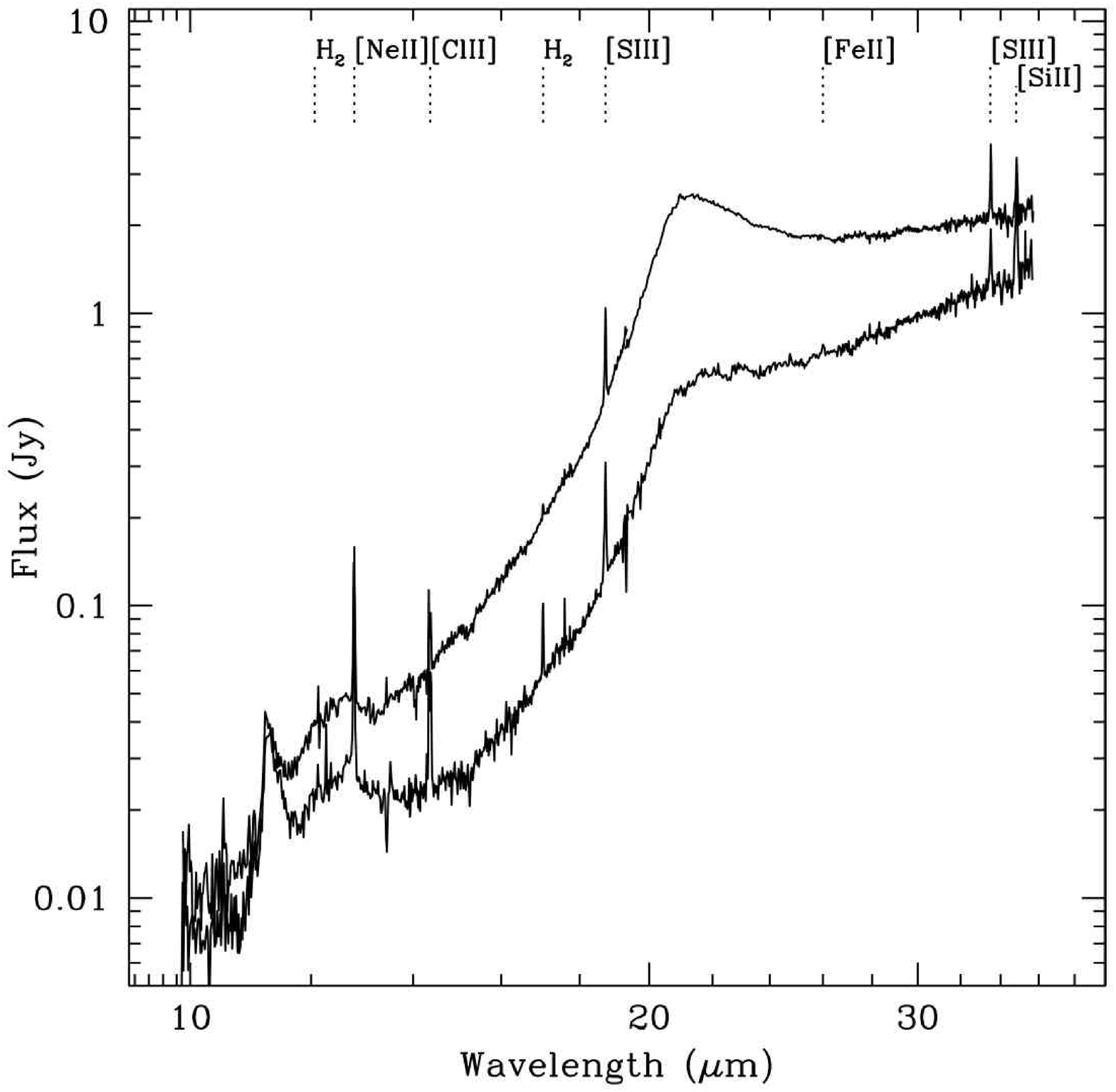} \caption{\label{highres}High resolution Spitzer IRS SH and LH spectra of the IR shell in G54.1+0.3. The fainter spectrum is from position 1, the interface between the PWN and the shell,  and the brighter spectrum is from position 2 at the IR knot (see Figure \ref{slits}).}
\end{figure*}


\begin{figure*}
\epsscale{1.0} \plotone{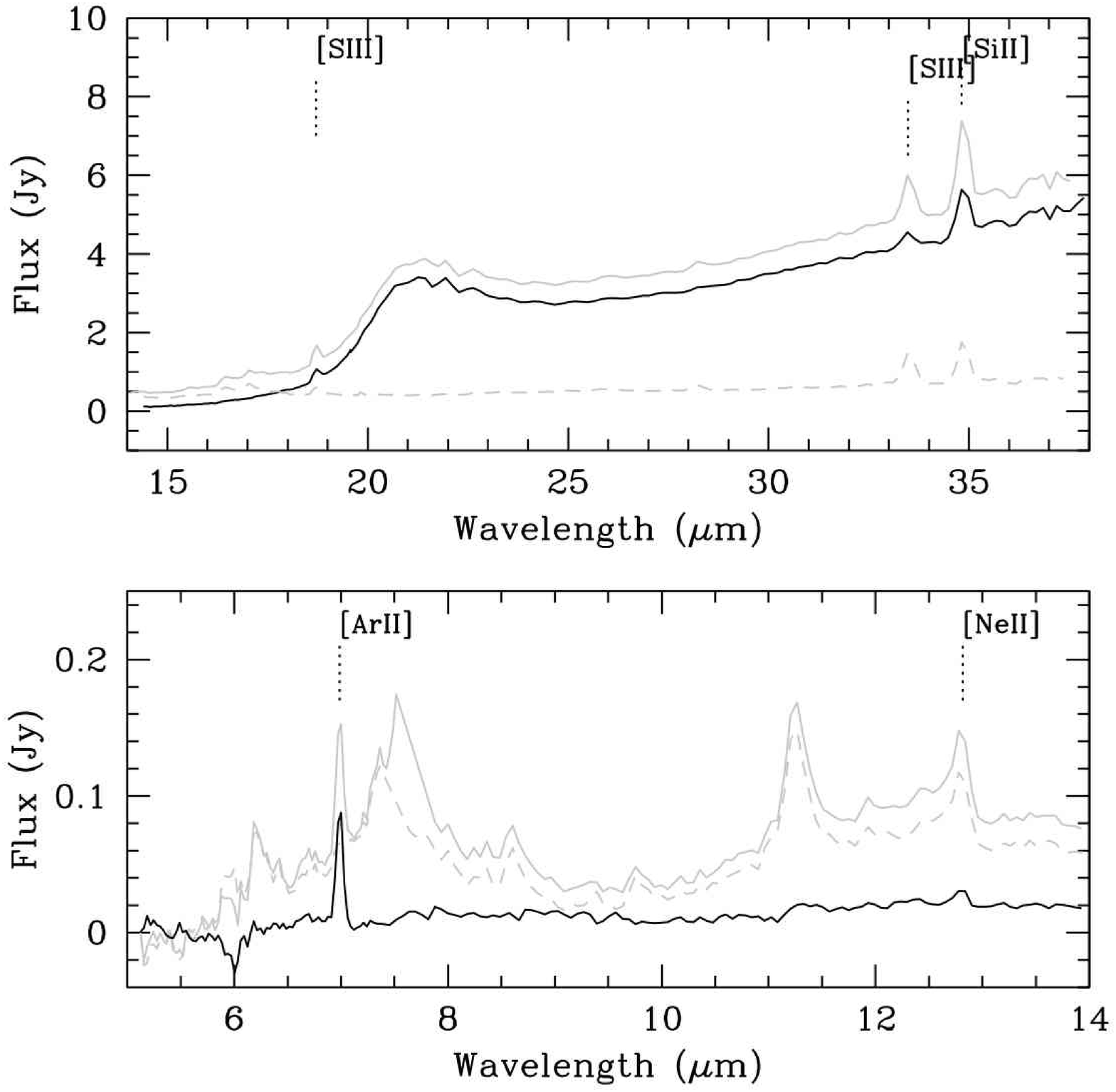} \caption{\label{lowres}Low resolution Spitzer IRS spectra extracted from the entire LL slit (top panel) and SL slit (bottom panel). The solid grey spectra represent spectra before background subtraction, dashed grey spectra are from the background extracted from the off-source nod positions, and the black spectra are the background-subtracted source spectra. The peak at around 7.5 \micron\ is due to an order mismatch at the edges, and was trimmed in the final spectrum.}
\end{figure*}

Curiously, the intensity of the [\ion{Si}{2}] line has a large peak at the apparent interface between the PWN and the IR shell. This peak is a factor of two brighter than the [\ion{Si}{2}] emission across the rest of the shell and does not correlate with spatial features shown by the IR continuum emission or with the [\ion{S}{3}] lines.  The relative intensities of the [\ion{S}{3}] and [\ion{Si}{2}] across the LL slit are also summarized in Table \ref{llspectab}, where the line fluxes were measured at 7 positions from spectra extracted from 3$\times$2 pixel sub-slits. The highest line broadening is seen at roughly the same position where the [\ion{Si}{2}] line peaks, position 1 of the high-resolution spectrum. As shown in Figure \ref{spatial}, the intensity of [\ion{Ar}{2}] as a function of position along the SL slit also peaks near the same position. Figure \ref{spatial2} shows the spatial profiles of the 24 \micron\ shell emission and the [\ion{Ar}{2}]  and [\ion{Ne}{2}] lines across the SL slit. The intensities of 24 \micron\ emission and [\ion{Ar}{2}]  appear to be anti-correlated spatially. The [\ion{Ar}{2}]  and [\ion{Ne}{2}] lines also appear to be anti-correlated across the SL slit. The Ne line is brightest along the northern part of the IR shell, positions 0--13 in Figure \ref{spatial2}. Figure \ref{spatial3} shows sample spectra from three different positions across the SL slit, extracted from 2 by 6 pixel sub-slits. The three positions show spectra with weak [\ion{Ne}{2}] emission, strong [\ion{Ne}{2}] emission, and emission with prominent broad features at 9 and 12 \micron. The spectrum with the strong broad features was extracted from the position of an embedded point source, suggesting that these features may be associated with the source, or enhanced in its vicinity.

\subsection{Evidence for Multiple Emission Regions}\label{components}


The IRS spectrum shows that the [\ion{Cl}{2}], [\ion{Fe}{2}], and [\ion{Si}{2}] lines are all broadened by a similar amount at both positions of the high-resolution slits. Their FWHM, deconvolved by the instrumental resolution (true FWHM in Table \ref{spitzertab}), is on the order of 1000 km/s. Another similarity is that these lines all have low ionization potentials of less than 13.6 eV.
The [\ion{S}{3}] lines show less broadening, have centroids that are slightly redshifted relative to the other observed emission lines, and have a higher ionization potential of 23.34 eV. The intensities of the low ionization lines do not appear to correlate spatially with the intensities of S and Ne lines. While the Si emission strongly peaks at the position where the PWN  appears to encounter the IR shell, the spatial plot of the S and Ne intensities shows no enhancement at this position. We are not able to determine if the S and Ne lines are spatially correlated, since the SL and LL slits sample different regions of the shell. The spatial distribution of the Ar line is anti-correlated with Ne, and it appears to peak at approximately the same region where Si is enhanced. It may be that the spectral line emission from the IR shell arises from distinct regions; a region where the highly broadened lines of ions with low ionization potentials are produced, including Si, Cl, Fe and perhaps Ar, a region from which most of the sulfur emission originates, and a region giving rise to the [\ion{Ne}{2}] line that may or may not be spatially correlated with sulfur. 
Similar components were observed by \citet{neu07}, who analyzed IRS spectral line maps from several SNRs interacting with molecular clouds, and found that the line emission generally originates from spatially distinct groups. The groups included lines from ions with low ionization potentials such as [\ion{Fe}{2}], [\ion{Si}{2}], and [\ion{P}{2}], lines of ions with ionization potentials higher than 13.6 eV, not including [\ion{S}{3}], and the lines of [\ion{S}{3}]. While the lines in various groups sometimes showed similar spatial distributions, the lines within a group were always strongly correlated \citep{neu07}. 
The emitting regions in G54.1+0.3 may represent distinct spatial components of different temperatures, or arise due to different critical densities of the ionized species. A possible explanation for the origin of different emission regions will be discussed in Section \ref{emissionregions}.


\begin{figure}
\epsscale{1.09} \plotone{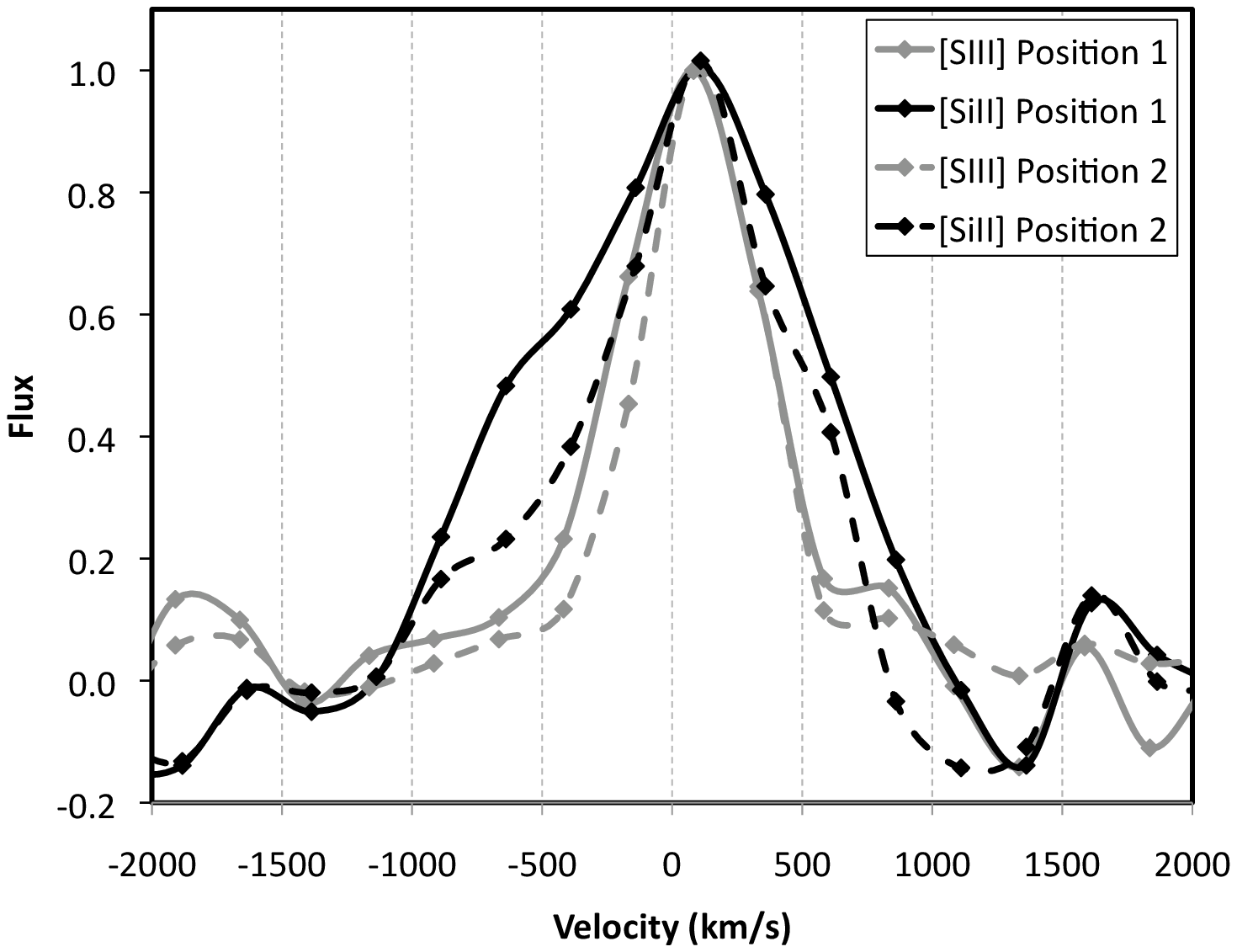} \caption{High resolution spectral profiles of the [\ion{Si}{2}] and [\ion{S}{3}]  33.5 \micron\ lines at two positions in the IR shell. The 1-$\sigma$ statistical uncertainties on each data point are on the order of 20\%. \label{profiles}}
\end{figure}

\subsection{Electron Density}\label{density}

\begin{figure*}
\epsscale{1.2} \plotone{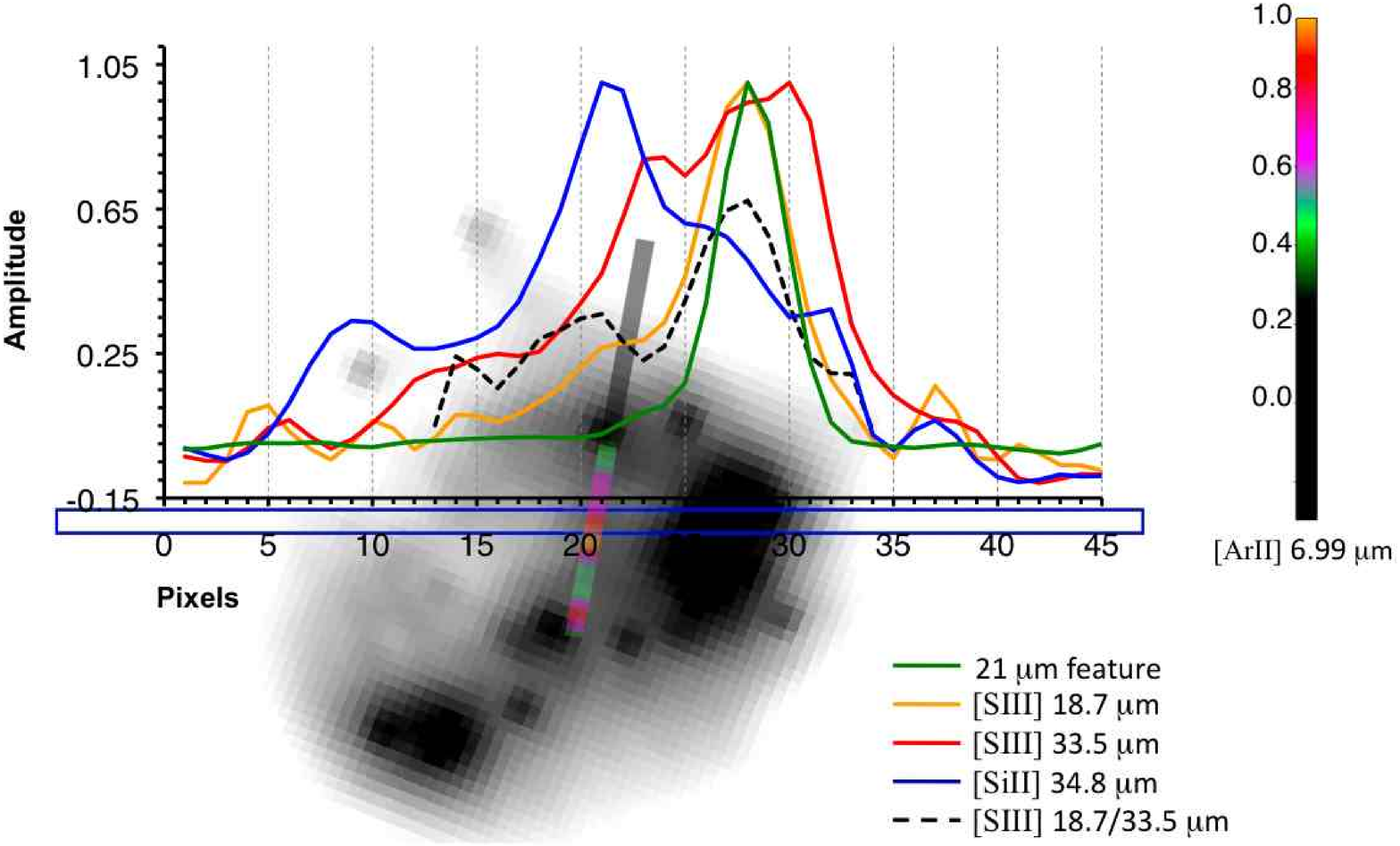} \caption{Intensity of various emission features in the G54.1+0.3 spectrum as a function of the position across the low-resolution IRS slits. The position of the LL slit, represented by the blue box, is overlaid on the MIPS 24 \micron\ image of G54.1+0.3. The pixel scale is 5.1\arcsec/pixel. The curves in the plot show the spatial intensity variation of the 21 \micron\ feature (green), and the [\ion{Si}{2}] (blue), [\ion{S}{3}]  (red \& orange) lines. The peak intensity of each feature has been normalized to unity and the amplitudes do not represent actual flux values. The black dotted line shows the actual ratio of the fluxes of the [\ion{S}{3}] 18.7 \micron\ and 33.5 \micron\ lines. The overlaid slit running N/S represents a normalized surface brightness map of the [\ion{Ar}{2}] line at 6.99 \micron. The statistical 1-$\sigma$ uncertainties of the intensities at each pixel are approximately 0.05 for the 21 \micron\ feature, 0.09 for [\ion{Si}{2}], 0.09 for [\ion{Ar}{2}], and 0.12 for [\ion{S}{3}] lines. \label{spatial}}
\end{figure*}

\begin{figure}
\epsscale{1.2} \plotone{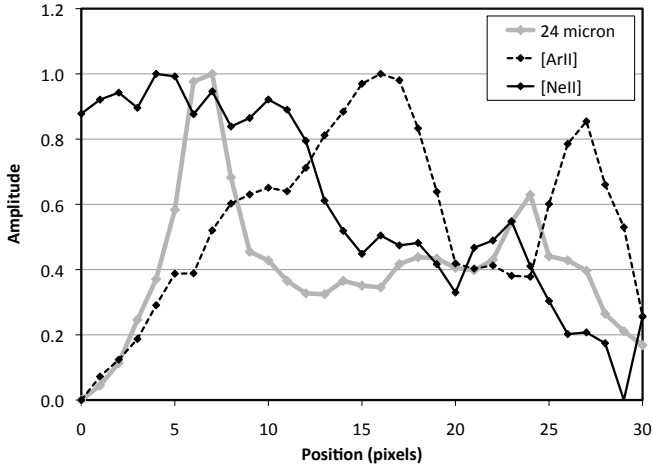} \caption{Intensity of the  [\ion{Ar}{2}],  [\ion{Ne}{2}], and 24 \micron\ shell emission as a function of position across the SL slit. The slit is shown in Fig. \ref{spatial}. Positions 0--30 correspond to positions across the SL slit from northern to southern edge of the IR shell emission. The statistical 1-$\sigma$ uncertainties on each data point are on the order of 15\%. The pixel scale for the SL slit is 1.8\arcsec/pixel. \label{spatial2}}
\end{figure}

The electron density of the emitting material can be probed using the temperature-insensitive density diagnostic ratio of the ground state fine structure lines of [\ion{S}{3}] at 18.7 \micron\ and 33.5 \micron. Both lines are present in the LL module, and the ratio of their surface brightness maps is shown by the dashed curve in Figure \ref{spatial}. Since the spectral maps were used to calculate the [\ion{S}{3}] line ratios across the slit, the absolute values of the point-to-point ratios in Figure \ref{spatial} are highly uncertain. Since the line profiles were not fitted in this case, the shape of the underlying continuum and the variation of the [\ion{S}{3}] line strength in the local background all contribute to the uncertainty. However, it is clear that the ratio peaks at the location of the bright IR knot, implying that the electron density here sharply increases. We estimated the electron densities using the extinction corrected intensities of the fitted [\ion{S}{3}] lines in the LH and SH spectra (Table \ref{spitzertab}) and at multiple positions across the LL slit (Table \ref{llspectab}). The calculation was performed using the IRAF STSDAS \textit {Nebular} package \citep{sha95}. The high resolution [\ion{S}{3}] 18.7 \micron/33.5 \micron\  line ratios are 0.7 $\pm$ 0.1 and 1.0 $\pm$ 0.1 for positions 1 and 2, respectively, giving a density range of $<$ 750 ${\rm\ cm^{-3}}$ at position 1, and 500--1300 ${\rm\ cm^{-3}}$ at position 2, for an assumed temperature range of 1000--10000 K. The quoted uncertainties are statistical uncertainties from the fit only. The ratios have additional uncertainties due to the fact that the LH and SH slits have different sizes and orientations, which affects the average surface brightness in the slit. Since the local background also contains [\ion{S}{3}] lines, spatial variations in the background add additional unquantifiable uncertainties to the line ratios and the density estimates. Table \ref{llspectab} shows the [\ion{S}{3}] ratio at six different positions across the LL slit. The ratios for all regions are consistent with being at the low density limit, except for the region corresponding to the IR knot at which the density is obviously higher, in the 500--2400 ${\rm\ cm^{-3}}$ range.

\subsection{Shock Diagnostics} \label{shock}

The PWN in G54.1+0.3  appears to fill the cavity of the IR shell, with the pulsar jet terminating at the IR knot (Figure \ref{3color}). Due to a morphological correlation between the PWN and the IR shell, it is possible and even likely that the pulsar wind is sweeping through the shell material.
A similar interaction is observed in the Crab Nebula, in which the PWN drives shocks into the SN ejecta and produces optical emission from a shell at the outer edge of the PWN \citep{san97}.
In order to estimate the velocity of the shock propagating into the shell material, we compared the measured spectral line intensities with
predictions from models of shock waves in SN ejecta.  Our model
code does not include chlorine.  However, Cl and Ar should be found
in the same region in the ejecta, and the excitation rates of the
lines are similar, so the ratio is mainly determined by the
abundance ratio. In addition, some emission may also originate in unshocked ejecta, and these ejecta may emit more efficiently in [\ion{Cl}{2}] than in [\ion{Ar}{2}] (see Section \ref{emissionregions}), as Cl$^+$ is more likely than Ar$^{+}$ to be
the dominant ionization species in freely-expanding unshocked ejecta.
 In the overlap region (Table \ref{lowspectab}), the [\ion{Ar}{2}]
line is 23 times as bright as [\ion{Cl}{2}], implying an abundance
ratio similar to \citet{woo95} nucleosynthesis models for core-collapse SNe.
We use the [\ion{Ar}{2}]/[\ion{Cl}{2}] ratio in the low-resolution to estimate the [\ion{Ar}{2}] intensities at Positions 1 and 2 of the high-resolution spectrum and scale
all the intensities to [\ion{Si}{2}] = 100 for ease of comparison in
Table \ref{shockmodels}.


\begin{deluxetable*}{crrrrrrrrr}
\tablecolumns{10} \tablewidth{0pc} \tablecaption{\label{shockmodels}SHOCK MODEL LINE INTENSITIES}
\tablehead{
\colhead{Shock Speed} &  \colhead{Ne II} & \colhead{Ne III} & \colhead{Si II} & \colhead{S III} & \colhead{S III} & \colhead{S IV} & \colhead{Ar II}  & \colhead{Ar III} & \colhead{Fe II} \\
\colhead{(km/s)} & \colhead{}  & \colhead{} & \colhead{} &
\colhead{(18 $\micron$)} & \colhead{(33 $\micron$)} & \colhead{} &
\colhead{} & \colhead{} & \colhead{} 
}
\startdata
\sidehead{\textbf{Model Intensities}}
20 & 8 & 0 & 100 & 5 & 5 & 0 & 44 & 1 & 5 \\
25 & 24 & 9 & 100 & 47 & 43  & 2 & 80 & 18 & 8 \\
30 & 42 & 101 & 100 & 194 & 150 & 32 & 118 & 115 & 11 \\
\sidehead{\textbf{Observed Intensities}}
Position 1 & 18 & $<$ 2 & 100 & 26 & 35 & $<$ 2 & 86 & $<$ 2 & 7 \\
Position 2 & 20 & $<$ 2 & 100 & 83 & 84 & 3 & 74 & $<$ 2 & $<$ 2 \\

\enddata
\tablecomments{Shock models were run using the code described in \citet{bla00} and \citet{wil08} and abundaces $H: He: C: N: O: Ne: Mg: Si: S: Ar: Ca: Fe$ = $12.0: 10.9: 10.0: 10.0: 14.0: 14.2: 14.3: 14.3: 14.0: 14.2: 13.0: 13.3$. For comparison with models, we normalized the observed intensities to [\ion{Si}{2}] 35 $\mu$m = 100. The [\ion{Ar}{2}] line is not measured at Positions 1 or 2, so we assume that the [\ion{Ar}{2}]$/$[\ion{Cl}{2}] ratio is constant. The preshock density of the models is 30 $\rm cm^{-3}$.}
\end{deluxetable*}


Models of emission from shocks in SN ejecta were run using the code
described in \citet{bla00} and \citet{wil08}. The results are summarized in Table \ref{shockmodels}. These models differ from models of shocks with normal astrophysical abundances
in that the enormous cooling rates generally give $T_e$ much lower
than $T_i$ in most of the cooling zone.
The models assume the abundance set
$   H:    He:    C:    N:    O:   Ne:   Mg:   Si:   S:    Ar:   Ca:   Fe$
= $12.0:  10.9: 10.0: 10.0: 14.0: 14.2: 14.3: 14.3: 14.0: 14.2: 13.0: 13.3$, given on a logarithmic scale,
but only the abundances of Ne, Si, S, Ar and Fe can really
be constrained by the observations.  The pre-shock density is
taken to be 30 $\rm cm^{-3}$, as suggested by the measured density of the IR knot. The density in the rest of the shell is somewhat lower, but  this should only affect the density sensitive [\ion{S}{3}] ratio.
The gas is followed until it cools to
300 K, and the pre-shock ionization state is taken to be
in equilibrium with the radiation field produced by the shock.

The models should not be considered unique, in that the pre-shock
ionization state, the abundances of coolants that
are not constrained by the observations (especially O), and the low
temperature cutoff of the models will all affect the predicted
spectra.  Moreover, it is unlikely that a single shock speed really
accounts for all the emission.  Therefore, we can draw only the rough
conclusions that
the intensity ratios (or limits on ratios) of [\ion{Ne}{3}]$/$[\ion{Ne}{2}], [\ion{S}{4}]$/$[\ion{S}{3}] and [\ion{Ar}{3}]$/$[\ion{Ar}{2}] suggest a shock speed of about 25 $\rm km\:s^{-1}$ or a little less.  The abundances of Ne, Si, S and Ar are comparable, while
that of Fe is about 10 times smaller.  We do not really have a
good handle on the O abundance, because the shock speed is too
low to produce much [\ion{O}{4}]. Overall, the emission lines indicate a shock moving through the
ejecta of a fairly massive SN, though a more O-rich composition
cannot be excluded. The high pre-shock density in the IR knot, and possibly in the rest of the shell,  suggests that shocks are driven into denser-than-average ejecta clumps (see Section \ref{sweptupejecta}), similar to what has been inferred for SNR B0540-69.3 \citep{wil08}.

\section{DUST EMISSION} \label{dustemission}

IR imaging and spectroscopy of G54.1+0.3 can provide important information about the properties and distribution of dust in the IR shell. The study of ISM dust provides constraints on models describing dust composition and properties. These models are best described by a grain composition that predominantly consists of astronomical silicates and carbonaceous material, such as PAHs, graphite, and amorphous carbon \citep[for a review see][]{dra09}. While a large fraction of interstellar dust is grown in the ISM \citep{dra09}, the rest is injected by stellar winds, PNe, novae, and SNe. Dust is formed in the cooling gas, where the density is high enough for grain growth, and the dust composition depends on the available gas abundances and environmental conditions in these objects. Models show that SN ejecta can have a range of compositions, with some of the most abundant grains being those of MgSiO$_3$, SiO$_2$, Mg$_2$SiO$_4$, Si, and C \citep[for a review see][]{koz09}. The shape of the IR spectrum and the presence of various emission features are clues that can help determine the mass and composition of dust in G54.1+0.3.

\subsection{The Unidentified 21 $\micron$ Feature}\label{21micron}

The IR spectrum of the shell (Figure \ref{highres}) shows a sharply rising continuum that flattens out beyond 21 \micron. The most obvious difference in the high resolution spectra at the two positions in the shell is the strength of the broad emission feature at 21 \micron. The feature is much more pronounced in the IR knot than in the rest of the shell. This is best seen in Figure \ref{spatial}, where the relative intensity of the emission in the 20.9--21.9 \micron\ range is shown as a function of the position along the LL slit. The precise shape of the spatial intensity plot is affected by the shape of the underlying continuum, but it is obvious that the peak intensity of the 21 \micron\  feature is an order of magnitude brighter at the IR knot. The feature peaks at approximately 20.8 \micron\ and has a FWHM of $\sim$ 3.7 \micron, although this estimate is sensitive to the determination of the underlying continuum emission. The shape is asymmetrical; it rises steeply at shorter wavelengths, and falls off more gradually longward of 20.8 \micron. Additional features observed in the IR spectra that may or may not be associated with the IR shell are emission features at $\sim$ 8.6 \micron\ and $\sim$ 12.4 \micron, with a FWHM of approximately 1.4 \micron\ and 2.1 \micron, respectively. These features are evident in Figure \ref{highres} and Figure \ref{spatial3}.

The 21 \micron\ feature in the IR shell of G54.1+0.3 is remarkably similar to the emission feature observed in the IR spectrum of Cas A \citep{rho08}. When we overlay our high-resolution spectrum from position 1 with the IRS spectrum of Cas A shown in Figure 3 of \citet{rho08}, we find that the features peak at the same wavelength and have the same width, implying that they likely arise from the same dust composition. In addition, the spectrum of Cas A also shows the broad emission features at 9 \micron\ and 12 \micron\ that are also evident in our spectrum, suggesting that these emission features may be correlated with the 21 \micron\ feature and produced by the same grain species. \citet{rho08} attributed the dust that produces the 21 \micron\ feature to freshly formed dust in the SN ejecta, and as we will discuss in Section \ref{interp}, we believe that the IR emission in G54.1+0.3 also originates from freshly-formed SN dust.

The IR spectrum of Cas A was fitted with multiple grain compositions, with the main contribution from SiO$_{2}$, Mg protosilicates, and FeO \citep{rho09}. While the overall continuum was well fitted by this composition, the fit had significant residuals from the 21 \micron\ feature and from the entire broad emission feature at 12 \micron.  The SiO$_2$ grains modeled as a continuous distribution of ellipsoids (CDE) provided a somewhat better fit to the 21 \micron\ feature in Cas A, in addition to being able to produce the 9 \micron\ feature when modeled at a higher temperature \citep{rho09}. In our attempts to fit the 21 \micron\ feature in the spectrum of G54.1+0.3, we found that the peak produced by the SiO$_2$ CDE composition was too narrow. However, if the SiO$_2$ grains are porous, the feature may be broadened and shifted to better match the observed profile. For this reason, these grains still may be plausible candidates for the production of the 21 \micron\ feature in Cas A and G54.1+0.3.

In order to gain insight into other possible grain species that could give rise to the 21 \micron\ emission, we turned to the extensive work that has been done in identifying the ``21 \micron\ feature" in carbon-rich protoplanetary nebulae (PPNe). The emission was discovered by \citet{kwo89} from IRAS observations of PPNe, and has since been observed in other dusty environments, including planetary nebulae with Wolf Rayet progenitors and two extreme carbon stars \citep[for a review see][]{pos04}. The emission feature observed in these objects actually peaks at a wavelength of 20.1 \micron, somewhat lower than the feature in G54.1+0.3. In a recent \textit{Spitzer} study of carbon-rich PPNe, high resolution spectroscopy revealed that the 21 \micron\ emission is accompanied by a narrow emission feature at 11.3 \micron\ and a broader feature at 12.3 \micron\ \citep{hri09}, similar to the feature in Cas A and what we observe in the IR shell. This provides some evidence that the features observed at various bands are related and that perhaps the same species give rise to emission in PPNe and in the IR shell of G54.1+0.3.

While various grain species have been proposed as the carriers of the 21 \micron\ resonance feature, including TiC, SiS$_2$, SiO$_2$, SiC, and FeO, all but the latter two have been discarded based on abundance constraints and the presence of other unobserved resonance bands \citep[e.g.,][]{zha09, spe04}. In a recent paper, \citet{zha09} ruled out all the mentioned carrier candidates, except FeO, due to an insufficient abundance of Ti, S, or Si in sixteen PPNe. They found that the FeO nano dust can reproduce the 21 \micron\ feature in these sources without producing unobserved secondary features nor exceeding the Fe budget. While FeO is a good fit to the 21 \micron\ feature in PPNe, the peak wavelength of 19.9 \micron\ falls short of the 20.8 \micron\ peak observed in G54.1+0.3 and Cas A. When modeled with FeO grains with a CDE distribution, the peak wavelength of the feature shifts to 20.7 \micron, but the FWHM increases to 6.5 \micron, far too broad to fit the feature we observe. FeO also fails to produce the other observed emission features at 9 \micron\ and 12 \micron\ that are likely related to the 21 \micron\ feature. 

\citet{spe04} suggested that the 21 \micron\ feature may be produced by SiC grains that contain impurities, and through laboratory experiments showed that doped SiC grains do indeed produce a resonance at 21 \micron\ that matches the shape of the feature in PPNe. They also found that nano-SiC grains produce a peak that is shifted redward by $\sim$ 1 \micron, matching the peak of the feature observed in the IR shell of G54.1+0.3 and Cas A. \citet{spe05} found that amorphous and nanocrystalline samples all produce three peak complexes near 9, 11, and 21 \micron. The SiC grains seem capable of producing all dust emission features that we observe in the shell of G54.1+0.3, including the sharp emission feature at 11.3 \micron, the broader emission features around 9 and 12 \micron, and the 21 \micron\ feature with the same peak wavelength as we observe. The relative strength of these features depends on the grain size and shape, and the temperature of the underlying continuum. The strength of the 21 \micron\ feature in particular, likely depends on the amount of carbon impurities in the SiC sample. 

Since strong emission from the 21 \micron\ feature is mostly observed in the PPN phase of stellar evolution, \citet{spe05} suggest a scenario in which the appearance of this feature is due to the particular physical conditions in these objects, rather than a production of new dust species. The 21 \micron\ emission would be suppressed in regions where the smallest grains have been destroyed by the central star, and in regions of the dust shell where the temperature and density are too low. In this case, the best candidates for the production of the emission feature would be cool ($\sim$ 100 K), micron sized $\beta$--SiC and nano-SiC grains that contain carbon impurities \citep[and references therein]{spe05}. Here, the $\alpha$-SiC and $\beta$-SiC refer to different polymorphs of silicon carbide grains.  

The scenario outlined in \citet{spe05} appears to be consistent with our observations of the IR shell in G54.1+0.3. The 21 \micron\ feature in G54.1+0.3 is strongly pronounced in the densest region, i.e. the IR knot, while it is suppressed in the diffuse shell emission that has a lower density. Our observations seem to suggest that SiC grains are strong candidates for producing the observed emission features. However, based on the models of \citet{koz09} that predict relative abundances of grain species formed in SN ejecta, SiO$_2$ is also a likely candidate. We cannot currently rule out either of these grain species. In order to identify the composition that produces the 21 \micron\ feature in G54.1+0.3, more detailed modeling of the IR spectrum is required. The dust in G54.1+0.3 is likely freshly-formed dust in the SN ejecta, with a similar composition as in Cas A.

\subsection{Extended IR Knot}\label{irknot}

\begin{figure}
\epsscale{1.2} \plotone{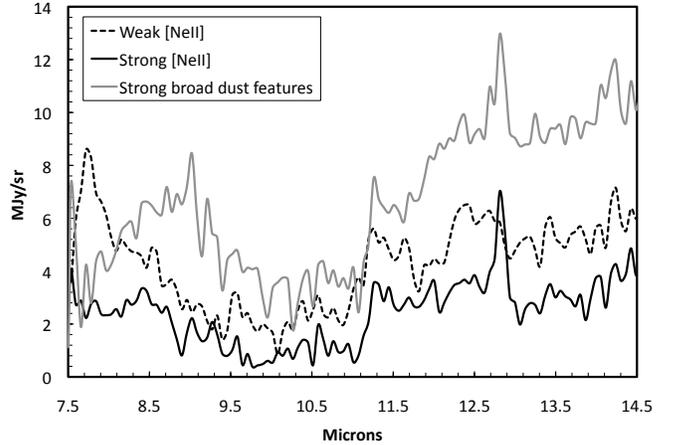} \caption{Sample spectra from three different positions along the SL slit. The spectra were extracted from 6 pixel long sub-slits at positions showing strong and weak [\ion{Ne}{2}] emission, and at the position coincident with an embedded point source, where the 9 and 12 \micron\ broad emission features are most prominent. The pixel scale for the SL slit is 1.8\arcsec/pixel. \label{spatial3}}
\end{figure}

The bright IR knot in the western part of the IR shell is shown in Figure \ref{irac}c, and in Figure \ref{3color} where the 24 \micron\ emission is shown in blue. The emission across the knot is not uniform, but appears to be concentrated in three separate clumps of emission whose spatial profiles are broader than the MIPS 24 \micron\ PSF. The extended clumps do not have counterparts at any of the IRAC wavelengths, nor does their emission correlate with the peak brightness of the MIPS 70 \micron\ image. The high-resolution IRS spectrum of the knot is dominated by a broad emission feature around 21 \micron\ and a rising underlying continuum (Figure \ref{highres}). The plot in Figure \ref{spatial} that depicts the spatial variation of emission features across the LL slit shows that the emission from the 21 \micron\ feature is an order of magnitude brighter at the IR knot than in the rest of the shell. We also note that the 9 and 12 \micron\ features are enhanced at the position of one of the point sources in the SL spectrum (Figure \ref{spatial3}). If these features correlate with the 21 \micron\ feature, then the 21 \micron\ feature may also be enhanced in the vicinity of the 11 point sources due to a higher temperature, thus causing the unusual [8]--[24] \micron\ color excess in the color--color plots of \citet{koo08}.

Since photoionization by the PWN cannot adequately heat the dust in the IR knot, we explore the possibility that this dust is shock-heated.
The pulsar's X-ray jet appears to terminate at the location of the knot (see Figure \ref{3color}), which may be an indication that it is driving a shock into the IR shell and compressing the dust to produce a small dense region with enhanced 21 \micron\ emission. The [\ion{S}{3}] 18.7 \micron\ emission and the ratio of the [\ion{S}{3}] 18.7 \micron\ to 33.5 \micron\ lines show a peak along the knot, implying a significant density enhancement in this region. The density calculated from the line ratios in the LH and SH at this position is in the 500--1300 ${\rm\ cm^{-3}}$ range. In order to check if this scenario is plausible, we estimated the fraction of the pulsar's spin-down luminosity ($\rm \dot E$) that would have to power the jet in order to get enough ram pressure to compress the dust in the shell. The spin-down luminosity $\rm \dot E$ for the pulsar in G54.1+0.3 is $\rm 1.2 \times 10^{37} \: erg \: s^{-1}$. 
Assuming a a shock velocity of 25 $\rm km \: s^{-1}$ (Section \ref{shock}), a preshock density of 30 $\rm cm^{-3}$, and a mean particle weight of $\rm 30\: m_p$, the
shock dissipates 1/2 $\rho v^3~=~0.012~\rm erg~cm^{-2}~s^{-1}$.
By estimating the area of the IR knot to be 600 square arcseconds (0.52 $\rm pc^2$ at a distance of 6 kpc), we calculate that the power in the jet would need to be $\rm \sim 6\times10^{34}\: erg \:s^{-1}$, or approximately half a percent of $\rm \dot E$, to produce the required ram pressure. This represents a reasonable (if approximate) fraction of the available spin-down power, indicating that the jet is compressing the dust in the shell and producing the denser IR knot with the pronounced 21 \micron\ feature.



\subsection{Dust Mass Estimate} \label{dustmass}

Since the spectrum of the IR shell is dominated by the emission from the 21 \micron\ feature, it was difficult to obtain a satisfactory fit to the spectrum with any single grain composition. In order to make a rough estimate of the dust mass in the IR shell, we fit the spectrum of the underlying continuum with two different grain composition models; astronomical silicates \citep{lao93} and Mg$_2$SiO$_4$ \citep{jag03}. We excluded the spectral region from 18--25 \micron\ where the emission from the 21 \micron\ feature dominates. In the fitting, we used the high-resolution spectrum from the diffuse shell emission, position 1, where the 21 \micron\ feature is somewhat suppressed, as well as the MIPS 70 \micron\ data to help constrain the dust temperature. Before fitting, the spectrum was scaled so that the integrated flux over the MIPS 24 \micron\ bandpass is equal to the total extinction-corrected flux at 24 \micron\ of 40 Jy. The fit was not satisfactory since it left large residuals from the 12 and 21 \micron\ features, but it can still be used to make a rough estimate of the amount of silicates that are present in the IR shell.  Grains of astronomical silicates at a temperature of $\sim$ 68 K, and Mg$_2$SiO$_4$ at $\sim$ 63 K, can approximately match the continuum and the 70 \micron\ data point, excluding the 18-25 \micron\ region and broad emission features at shorter wavelengths. The total dust mass in the shell is given by

\begin{equation}
{M}_{dust}=\frac{{F}_{\nu}d^{2}}{{B}_{\nu}(T_{d})}\frac{4{\rho}a}{3Q_{abs}},
\end{equation}
where F$_{\nu}$ is the total IR flux, $\rm d$ is the distance, B$_{\nu}$ is the Planck function evaluated at the grain temperature, $\rm \rho$ and $\rm a$ are grain density and size, and $\rm Q_{abs}$ is the absorption efficiency. Absorption efficiencies for 0.05 \micron\ sized astronomical silicates and forsterite are from \citet{lao93} and \citet{jag03}, respectively. The assumed distance is 6 kpc. We find an approximate dust mass of 0.02 M$_{\odot}$ for astronomical silicates, and 0.04 M$_{\odot}$ for forsterite. Since the fits were not satisfactory for all of the observed emission features, these values should only serve as rough estimates. We note that this estimate is in the same range as the dust mass estimated for the Crab Nebula \citep{tem06}.

\section{PHYSICAL INTERPRETATION FOR THE IR SHELL}\label{interp}

\begin{figure}
\epsscale{1.2} \plotone{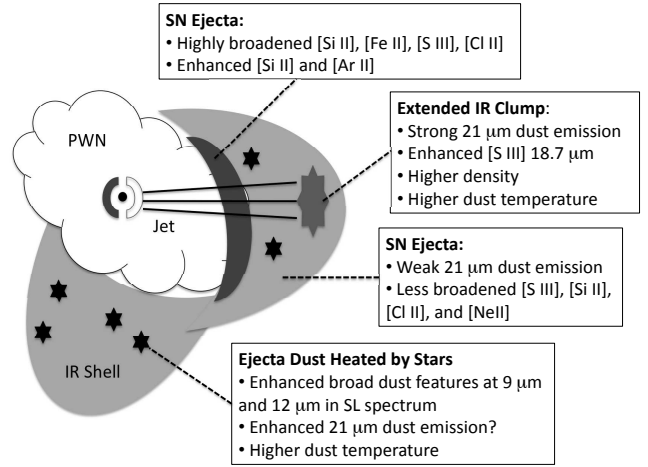} \caption{Summary of the IRS spectroscopic results for the IR shell in G54.1+0.3. \label{cartoon}}
\end{figure}

The cartoon in Figure \ref{cartoon} summarizes the IRS spectroscopic results for different regions of the IR shell surrounding the PWN in G54.1+0.3. The PWN fills the cavity of the IR shell, with the diffuse X-ray emission extending almost to the outer boundary of the MIPS 24 \micron\ emission (Figure \ref{3color}). The dark grey region in Figure \ref{cartoon} represents the location where we observe enhanced Si and Ar abundances, and where the intensity of the silicon line shows a sharp rise (see Figure \ref{spatial}.) The highest line broadening is also observed in this region (Table \ref{highres}). The diffuse shell emission, represented by light grey, shows lines that are somewhat less broadened and continuum emission with a weak 21 \micron\ dust feature. The pulsar's jet appears to terminate at the bright IR knot whose spectrum shows strong emission from the 21 \micron\ feature and a higher density, as derived from the \ion{S}{3} 18.7/33.5 \micron\ ratio. 


In this section, we consider two possible physical scenarios that may explain the observations of the IR shell. A previous explanation is that the shell formed by the progenitor's wind that compressed the surrounding medium and then fragmented and collapsed to form YSOs, as suggested by \citet{koo08}. An alternative scenario is that the shell is composed of expanding SN ejecta, and that the observed point sources are produced by radiative heating of ejecta dust by early-type stars embedded within the expanding SNR. Based on the IR data, and the lack of thermal X-ray emission from the shell, we find the ejecta interpretation a more plausible explanation for the origin of thermal dust emission in G54.1+0.3. 


\subsection{Preexisting Shell}
 
The 24 \micron\ MIPS image of the IR shell in G54.1+0.3 contains 11 point sources embedded in the diffuse shell emission that are arranged in a ring-like structure (Figure \ref{slits}). \citet{koo08} suggest that these point sources are massive pre-main-sequence stars, larger than 10 $M_{\odot}$, with an age of $\lesssim 2$ Myr. Their formation is suggested to be triggered by the progenitor star of G54.1+0.3 during its post-main-sequence stage of evolution. Most of the point sources are detected in IRAC and in the near-IR 2MASS images, and some even have optical counterparts. Their JHK$_s$ color-color and color-magnitude diagrams indicate that they are young stars with spectral types ranging from B1.5 to O8. 

\citet{koo08} show that the spectral energy distribution (SED) of the brightest star in the K$_s$ band, the southernmost point source in Figure \ref{slits}, is similar to the spectral energy distribution of early Herbig Be stars. The IRAC and MIPS 24 \micron\ colors, however, are not consistent with the usual colors of Class 0/I and Class II YSOs. While the stars in the IR loop have a small IRAC [3.6] - [5.8] excess, consistent with Class II sources, the [8] - [24] is much higher than expected, in the 6 to 8.5 magnitude range \citep{koo08}. The IR spectrum in Figure \ref{highres} shows that there is a broad emission feature peaking around 21 \micron. This feature likely dominates the MIPS 24 \micron\ flux throughout the diffuse shell, and it may be the cause of the unusually high [8.0] - [24] excess. It may be that the point source photometry is contaminated by the emission from the diffuse shell, but since the observed 24 \micron\ excess is up to 4 magnitudes higher than what is expected for Class II YSOs, it appears more likely that the emission from the 21 \micron\ feature is enhanced in the vicinity of the these sources. A similar enhancement is observed in the IR knot, discussed in Section \ref{irknot}. 

If the point sources in the IR shell are embedded YSOs, their presence would require that the shell was formed prior to the SN explosion. 
However, the high expansion velocity of over 500 $\rm km\:s^{-1}$, as derived from the silicon, iron, and chlorine lines, indicates that these lines have to come from rapidly expanding SN ejecta. 
Another inconsistency with the interpretation of a preexisting shell is the absence of thermal X-ray emission in G54.1+0.3, and a lack of evidence for any interaction with the supernova blast wave. If the SNR has encountered the shell, we would expect the reverse shock to have disrupted the PWN. The structure of the PWN shows no indication that the reverse shock has reached its surface, providing further evidence against this scenario. In the following section, we provide an alternative interpretation for the origin of the IR point sources and the diffuse shell emission, without the need for a preexisting shell.

\subsection{SN Ejecta} \label{sweptupejecta}

Here, we consider a scenario in which the IR shell is composed entirely of SN ejecta. The shock models imply that the PWN overtakes the freely-expanding ejecta with a shock velocity of $\sim$ 25 $\rm\ km\:s^{-1}$ (see Section \ref{shock}). The line emission from the IR knot, where the pulsar's jet compresses the ejecta, is likely dominated by shock emission. However, in the rest of the shell where we only have upper limits on the density, photoionization by the stellar radiation field and the PWN likely accounts for most of the emission. If the density of the shell is significantly lower than in the IR knot, the surface brightness of [\ion{Si}{2}] produced by slow shocks would be too low to match the observed [\ion{Si}{2}] intensity. However, if the density of the shell is comparable to the density of the IR knot, the ram pressure of the PWN would be too low to drive a 25 km/s shock into such dense ejecta. Emission from the shock most likely dominates only in the region where the jet drives a slow shock into the dense IR knot region, while the emission from photoionization dominates in the rest of the shell. The sharp [\ion{Si}{2}] peak at the inner edge of the shell (Figure \ref{spatial}) may originate from a region with a density that is lower than in the IR knot, but high relative to the rest of the shell.This region may be produced by the compression of gas by the PWN, but with a lower ram pressure than in the jet.

The high inferred preshock ejecta density in the IR knot implies that
dense ejecta clumps with a low volume filling fraction are embedded in
less dense interclump medium. Such clumps are expected to
be produced by compression of ejecta by a radioactively-heated and
expanding Ni bubble. A similar scenario has been suggested for SNR 0540-69.3 \citep{wil08}. From equation (5) in Williams et al. (2008) that is
valid for a Type IIP ejecta model, the predicted densities in a shell
swept up by the Ni bubble are $\rm \sim 250 (t_{SNR}/1500\: yr)^{-3}
(M_{ej}/15 M_\odot)^{5/2} E_{51}^{-3/2}\:amu\:cm^{-3}$. At 1500 yr and
with $M_{ej} = 20 M_\odot$, $E_{51}=1$, this gives 500 amu cm$^{-3}$. Assuming a mean particle mass of $\sim$ 30 $\rm m_p$, and a preshock clump density of $\rm \sim 30 \: cm^{-3}$, this is in reasonable agreement with observations.

The broadened line emission suggests an ejecta velocity greater than $\sim$ 500 $\rm\ km\:s^{-1}$, since we are observing regions away from the center of the remnant where the expansion is mostly transverse to the line of sight. The point sources observed at 24 \micron\ can be explained if we consider a scenario in which the SN exploded inside a cluster of young stars. Assuming a distance of 6 kpc, the observed stellar density in the region covering the IR shell is not unusual for a young cluster \citep[e.g.][]{elm00}. In this scenario, the SNR blast wave has expanded beyond some of the cluster stars, embedding them within the expanding SN ejecta. The ejecta dust is then blowing by the embedded stars and being radiatively heated. The dust model described in Section \ref{dustmodel} shows that if the dust is heated by a dozen such stars, this can reproduce the observed 24/70 \micron\ ratio and give rise to the IR point sources at 24 \micron\ without a need for triggered star formation. The fact that the spectrum of the shell closely resembles the spectrum of freshly-formed dust in Cas A \citep{rho08} provides additional support for this interpretation.

 
\subsubsection{Origin of Multiple Emission Regions} \label{emissionregions}

As discussed in the previous section, the line emission from the shell likely arises from a combination of slow shocks driven into dense ejecta and photoionization. The IR spectra do indeed show that different emission lines have different spatial and kinematic properties (Section \ref{spatialvariation}) and that they may arise from distinct components in the shell (Section \ref{components}). Intense UV stellar radiation within the stellar cluster may be heating the ejecta gas either by directly ionizing it or by ejecting photoelectrons from grains.
Ionization of H and O requires photons with energies greater than 13.6
eV, that may be scarce if only B-type stars were present within the
cluster. Abundant, less energetic UV photons that heat the dust also
eject photoelectrons. Ejecta gas heated by grain photoelectrons is
expected to cool by emission in IR lines of singly ionized species with
ionization potentials less than 13.6 eV, including [\ion{Si}{2}] 34.8
$\mu$m, [\ion{Fe}{2}] 26.0 $\mu$m, and [\ion{Cl}{2}] 14.4 $\mu$m.
Unlike Si and Fe, Cl is not expected to be strongly depleted onto dust,
and this minor ejecta constituent may be the major coolant in
freely-expanding ejecta. 

Heating by stellar photons should be effective both in the ejecta
shell swept up by the PWN and in the freely expanding ejecta ahead of the shell. At sufficiently large radial distances, free-expansion velocity must exceed the shell velocity. If heating
were still  effective there, widths of [\ion{Si}{2}], [\ion{Fe}{2}], and
[\ion{Cl}{2}] lines could be be larger than widths of lines such as
[\ion{S}{3}] that are more efficiently produced by shocks in the swept-up PWN shell.
This is what is observed (Table \ref{spitzertab}), suggesting that heating of ejecta
gas by grain photoelectrons, and perhaps also by direct photoionization
by more energetic photons, occurs in G54.1+0.3.

\subsubsection{Origin of IR Point Sources} \label{dustmodel}

Here, we describe a model in which the ejecta dust is being heated by stellar sources inside a cluster in which the SN exploded. The stellar radiation field decreases in intensity as $r^{-2}$ with increasing distance $r$ from a star (and steeper still if the dust
optical depth $\tau$ is non-negligible), so dust is heated to much
higher temperatures close to the star than far away from it. Hotter dust
re-radiates the absorbed stellar radiation at shorter wavelengths than
the more distant cooler dust, an effect that produces an apparent
infrared excess in the stellar spectrum. We demonstrate this effect with
a simple dust model, which is also used to estimate the total ejecta
dust mass in G54.1+0.3.

We consider a B0V star with a luminosity of 25,000 $L_\odot$ and
temperature of 30,000 K, surrounded by uniformly distributed ejecta
dust. There is a central cavity devoid of dust in the immediate vicinity
of the star, where dust is cleared either by a stellar wind or by the
stellar radiation pressure \citep{art97}, but this cavity is
expected to be small in view of the large ram pressure exerted by
fast-moving SN ejecta. We assume a small cavity radius of 0.003 pc;
properties of the mid-IR emission are insensitive to the assumed cavity
radius as long as the cavity remains small. Absorption of stellar UV
radiation by dust and its re-radiation as thermal dust emission in the IR
are modeled as described by \citet{bor94}. These dust models
include effects of grain temperature fluctuations that are important for
small grains heated by energetic UV photons. Scattering of photons by
dust is not included, so these models underestimate absorption if $\tau$
is not small. Multiple dust species are predicted to form in SN
ejecta \citep{koz09}, one of the most abundant being forsterite
(Mg$_2$SiO$_4$). We considered amorphous forsterite grains in our
models, with density of 3.3 g cm$^{-3}$ and optical constants from
\citet{sco96}. As a single grain size is unlikely, a power law
distribution in grain sizes was assumed, with a power-law index of
$-3.5$, ranging in radii from 0.001 $\mu$m to 0.25 $\mu$m. We chose
0.007 $M_\odot$ pc$^{-3}$ as the spaced-averaged grain mass density; within a factor of a few, this choice is compatible with the dust mass estimate presented in Section \ref{dustmass} and the volume occupied by the IR shell.


 \begin{figure}
\epsscale{1.2}
\plotone{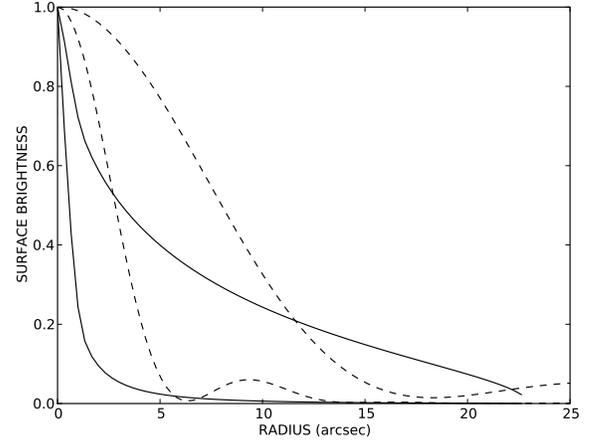}
\caption{Normalized model profiles in the 24 $\mu$m and 70 $\mu$m MIPS
bands (left and right solid lines, respectively). The corresponding {\it
Spitzer's} PSFs are plotted by dashed lines. Because of a large radial
gradient in dust temperature, IR emission is spatially extended at 70
$\mu$m, but a point-like emission enhancement appears at 24 $\mu$m and at
shorter IRAC wavelengths.
\label{spatialprofiles} }
\end{figure}

Our dust model matches the extinction-corrected 70/24 $\mu$m MIPS flux
ratio of 1.9 if we choose the radius to which the dust extends to be 0.69 pc. The 70/24 MIPS flux
ratio decreases outward, tracing a steep radial dust temperature
gradient, so models with a smaller (larger) radius under(over)predict
this ratio. We consider only dust within a radius of 0.69 pc in our
discussion of spatial IR profiles and of dust mass estimates. Within
this radius, the infrared dust luminosity is 1200 $L_\odot$, comprising
only 4.8\%\ of the stellar luminosity, so the SN ejecta are
optically thin to stellar radiation. The predicted surface brightness in
the 24 $\mu$m and 70 $\mu$m MIPS bands is shown in Figure
\ref{spatialprofiles}, without accounting for the limited spatial
resolution of {\it Spitzer}. A sharp, point-like emission enhancement is
present at 24 $\mu$m, and absent at 70 $\mu$m. A comparison with the
{\it Spitzer} PSF (Figure \ref{spatialprofiles}) reveals that the 24
$\mu$m emission enhancement at the center cannot be spatially resolved
with {\it Spitzer}. The uniformly-distributed ejecta dust appears as an
unresolved point source, surrounded by faint extended emission. This
is no longer true at 70 $\mu$m; emission at this wavelength is
predominantly produced far from the star, and {\it Spitzer's}
spatial resolution becomes adequate for mapping the spatially-extended
dust emission. This simple dust model matches well the observed
morphology of G54.1+0.3 as seen by the IRAC, MIPS and {\it Akari} detectors \citep{koo08}.

The 70 $\mu$m flux, equal to 7.5 Jy in our model with a radius of 0.69 pc,
is produced by 0.0097 $M_\odot$ of forsterite dust. Since the total
spatially-integrated 70 $\mu$m flux is equal to 76 Jy, the 11 stars
with an apparent infrared excess identified by \citet{koo08} are
sufficient to heat ejecta dust with an estimated total mass of approximately 0.1
$M_\odot$. This mass is within about a factor of two of the mass estimate from forsterite grains, derived from spectral fitting is Section \ref{dustmass}.
This dust mass estimate can be improved once we learn more 
about these stars from their near IR spectroscopy, and once the dust composition is better constrained. In particular, a
better understanding of the stellar radiation field in the parent
stellar cluster (or association) of the G54.1+0.3 progenitor will allow
for more realistic modeling of dust heating by stellar UV photons than
currently possible.

\section{CONCLUSIONS} \label{concl}

In this paper, we presented deep \chandra observations of G54.1+0.3 and \spitzer imaging and spectroscopy of the surrounding IR shell. The 300 ks \chandra observation was used to derive a new value for the absorbing column density toward G54.1+0.3 that was then used in the IR analysis. A more detailed X-ray study will be presented in a separate publication. We fitted the X-ray spectra from different regions of G54.1+0.3 with an absorbed power-law model and found that a more accurate value of N$_H$ results when the piled-up pulsar region is excluded from the fit. The new value of N$_H$ is (1.95$\pm$0.04)$\times10^{22}\rm\ cm^{-2}$, somewhat larger than the previous estimate by \citet{lu02} of (1.6$\pm$0.1)$\times10^{22} \rm\ cm^{-2}$.

The IRAC and MIPS images reveal an IR shell with a dozen point sources arranged in a ring-like structure, and a dense extended region that is aligned with the pulsar's jet.
IRS spectra reveal a number of ionic emission lines, including [\ion{Ar}{2}], [\ion{Ne}{2}],[\ion{Cl}{2}], [\ion{S}{3}], [\ion{Fe}{2}], and [\ion{Si}{2}], and a rising continuum with a broad emission feature at 21 \micron. 
We conclude that the emission from the IR shell likely arises from SN
ejecta and freshly-formed SN dust. The evidence for this
interpretation comes from the observed morphology of the PWN and the
shell, significant broadening of IR emission lines, comparisons of
line intensities with shock models, and the observed features in the
dust emission spectrum.
Since the PWN fits into the cavity of the IR shell, these components are almost certainly associated and likely interacting. The spectral lines show evidence for broadening with a maximum FWHM of over 1000 $\rm\ km\:s^{-1}$, suggesting that these lines arise in rapidly expanding SN ejecta. Based on our shock model results and density diagnostics, the observed line intensities are consistent with a scenario in which the emission from the IR knot is produced by the pulsar's jet driving a $\sim$ 25 $\rm\ km\:s^{-1}$ shock into a dense ejecta clump. The emission from the rest of the shell is most likely dominated by photoionization of the ejecta gas.  
%
The [\ion{S}{3}] 18.7/33.5 \micron\ line is consistent with a low
density limit in the IR shell and a higher density in the IR knot of
$\sim$ 1000 $\rm\ cm^{-3}$. This extended IR knot may be produced as
the pulsar's jet compresses the shell material. In order for the
jet to have enough ram pressure to produce the density enhancement, it
would need to be powered by approximately 0.5 \% of the pulsar's
spin-down luminosity $\dot E$. If the density in the rest of the IR shell is near the estimated upper limit, i.e. comparable to the density of the IR knot, then the shell would have to be composed of dense ejecta clumps with a low volume filling fraction, as produced by a radioactively heated and expanding Ni bubble.

The broad 21 \micron\ emission feature is enhanced at the position of the knot and weak in the rest of the shell. The profile of the feature is remarkably similar to the spectrum of Cas A \citep{rho08}, and is most likely  produced by the same dust species. The emission may be caused by SiO$_2$ grains, as suggested by \citet{rho09}, but SiC grains are also likely candidates since they are capable of producing the 21 \micron\ feature and other associated emission features around 9 and 12 \micron\ \citep{spe05}. We estimate the dust mass in the IR shell to be in the 0.04--0.1 M$_{\odot}$ range, assuming a forsterite grain composition. More detailed modeling of the IR spectrum is required to better characterize the properties and composition of the dust.

We find that our observations are best described by a scenario in which the shell emission arises entirely from SN ejecta. In this case, the point sources in the shell are not young stellar objects, but would be attributed to radiatively heated ejecta dust by early-type stars that belong to a stellar cluster in which the SN exploded. A simple dust model shows that this scenario can reproduce the observed IR emission, including the IR point sources. The stars are illuminating unshocked dust grains and we are probing freshly formed SN dust before its encounter with the reverse shock. The study of dust production in SNe may prove to be most effective in cases where the SN explosion occurs inside a young stellar cluster. 

\acknowledgments

POS acknowledges partial support from NASA contract NAS8-03060. TT acknowleges partial support from CXC grant GO8-9064X. We would like to thank Alexey Vikhlinin for the ACIS data reduction script, Joe Hora for useful discussion on the  21 \micron\ feature in PPNe, Achim Tappe for tips on CUBISM line intensity maps, Charles Lada for discussion on young stars, Anne Hofmeister for nano-SiC absorbance data, and Jeonghee Rho for the discussion on the 21 \micron\ feature in Cas A. This work is based in part on observations made with the Spitzer Space Telescope, which is operated by the Jet Propulsion Laboratory, California Institute of Technology under a contract with NASA. The IRS was a collaborative venture between Cornell University and Ball Aerospace Corporation funded by NASA through the Jet Propulsion Laboratory and Ames Research Center. SMART was developed by the IRS Team at Cornell University and is available through the Spitzer Science Center at Caltech.

\bibliographystyle{plain}










\end{document}